\documentclass[10pt,a4paper,twocolumn,aps,prd,notitlepage,nofootinbib,showpacs]{revtex4-1}
\usepackage[latin9]{inputenc}
\usepackage{amsmath}
\usepackage{graphicx}

\makeatletter

\pdfpageheight\paperheight
\pdfpagewidth\paperwidth

\usepackage{color}
\usepackage[normalem]{ulem}

\makeatother

\begin{document}

\title{Self-gravitating axially symmetric disks in general-relativistic
rotation}

\author{Janusz Karkowski}

\author{Wojciech Kulczycki}

\author{Patryk Mach}

\author{Edward Malec}

\author{Andrzej Odrzywo\l ek}

\author{Micha\l ~ Pir\'og}

\affiliation{Instytut Fizyki im.~Mariana Smoluchowskiego, Uniwersytet Jagiello\'{n}ski,
{\L }ojasiewicza 11, 30-348 Krak\'ow, Poland}
\begin{abstract}
We integrate numerically axially symmetric stationary Einstein equations
describing self-gravitating disks around spinless black holes. The
numerical scheme is based on a method developed by Shibata, but contains
important new ingredients. We derive a new general-relativistic Keplerian
rotation law for self-gravitating disks around spinning black holes.
Former results concerning rotation around spinless black holes emerge
in the limit of a vanishing spin parameter. These rotation curves
might be used for the description of rotating stars, after appropriate
modification around the symmetry axis. They can be applied to the
description of compact torus\textendash black hole configurations,
including active galactic nuclei or products of coalescences of two
neutron stars. 
\end{abstract}

\maketitle

\section{Introduction}

Stationary axially symmetric systems of rotating polytropic fluid
share two common properties both in general relativity and in Newtonian
gravity. They need additional specification\textemdash the so-called
rotation curve, that tells particles of fluid how to rotate. They
are free-boundary elliptic systems\textemdash the shape of rotating
fluids cannot be set \textit{a priori}, but constitutes a part of
the solution. In Newtonian gravity the Poincar\'{e}-Wavre theorem
allows one to define explicitly a rich family of allowed simple rotations
$\Omega=\Omega(\varpi)$, where $\Omega$ is the angular velocity,
and $\varpi$ denotes the distance from the rotation axis. In contrast
to that, the general-relativistic laws of rotation had been rather
poorly recognised. The classical rotation curves used in general relativity
\cite{Butterworth_Ipser,Bardeen_1970} are essentially variants of
uniform rotation. Keplerian rotations are common in rotating astrophysical
systems, but in Newtonian limits of known rotation curves \cite{Butterworth_Ipser,Bardeen_1970,GYE,UTGHSTY,Uryu}
the Keplerian rotation law is obtained only approximately.

We proposed in \cite{MM} a general-relativistic differential rotation
law $j=j(\Omega)$, where $j$ denotes the specific angular momentum,
capable to describe a stationary system consisting of a self-gravitating
disk circulating around a spinless (or spinning) black hole. Its most
striking characteristic is that the nonrelativistic limit exactly
coincides with Newtonian angular velocities $\Omega=w/\varpi^{\lambda}$
($0\le\lambda\le2$). In particular the Keplerian rotation law $\Omega\propto1/\varpi^{3/2}$
belongs to the Newtonian limit of this general-relativistic rotation
curve. The rotation law \cite{MM} has been recently investigated
up to the first order of the post-Newtonian approximation (1PN), for
a spinless black hole \cite{KMMPX}.

One of the two main aims of this paper is the generalization of results
obtained in \cite{MM}. We shall derive a general-relativistic Keplerian
rotation law, that might be more convenient for the description of
tori moving around \textit{spinning} black holes.

From the formal point of view the admissible rotation law can be specified
freely both in general relativity, as a function $j=j(\Omega)$, and
in Newtonian gravity, as $\Omega=\Omega(\varpi)$. However not for
every rotation law the coupled system of Einstein\textendash Euler
or Poisson\textendash Euler equations admits solutions for a reasonably
wide range of solution parameters. Since sufficiently general exact
existence results are at present only available for the Newtonian
case \cite{Auchmuty}, one has to resort to a numerical evidence of
existence or absence of solutions. As we will argue in the forthcoming
sections, this procedure actually yields practical information on
the form of robust rotation laws.

Our second main objective is to show that compact systems consisting
of a spinless black hole and a torus in the general-relativistic rotation
\cite{MM} (that depends on only one parameter) can be described numerically
in the full Einstein theory. We describe the black hole\textendash toroid
system within the puncture framework implemented in \cite{MSH}, but
there are important modifications. One of them is that adopted rotation
laws lead to a set of highly nonlinear algebraic equations that need
to be solved during iterations. We imposed in a different way than
Hachisu \cite{Hachisu} one of construction conditions for numerical
toroids, that significantly improved convergence of iterations. We
test our code on the Kerr solution, which in turn led to implementations
of some boundary conditions that might be different.

Recent investigation shows that there are reasons to conjecture that
stellar mass black hole\textendash torus systems emerge as end products
of binary neutron stars or black hole\textendash neutron star mergers
{[}\cite{BR2017} and references therein{]}, associated with some
detections of gravitational waves. Moreover, fluid tori in these systems
do \textit{self-gravitate and rotate with the Keplerian velocity } \cite{NSkeplerian,KR,ECGKK,SFHKKST}
around a spinning black hole.

The present paper gives a complete description of the formalism and
obtained results. In the next section we briefly describe the Einstein
equations and the equations of hydrodynamics, explaining the rotation
law of \cite{MM}. Relevant information on the numerical scheme is
given in Sec.\ III. Section IV reports results of numerical tests.
In particular, we perform standard convergence tests and recover numerical
results of Shibata \cite{MSH}. We discuss in Sec.\ V the status
of rotation parameters in the rotation law of \cite{MM}; it appears
that the rotation law is rather rigid and there is only one parameter
that is free and another that emerges as a part of the numerical solution.
In the subsequent section we show profiles of a few toroids for different
systems\textemdash with constant angular momentum density, the Keplerian
angular rotation and for systems with tori in (almost) uniform linear
velocity rotation. Section VII is dedicated to the derivation of a
new rotation curve that is expected to be efficient in describing
systems with spinning black holes. The result of \cite{MM} appears
in the limit of the vanishing spin of the black hole. Section VIII
shows explicitly that the Poincar\'{e}-Wavre (Newtonian) property
of the angular velocity\textemdash that the angular velocity does
not depend on the height above the symmetry plane\textemdash can be
robustly broken in general relativity. Finally, we summarize obtained
results and outline prospects for further related research.

We investigated the whole interval of $a/m\in[-0.9,0.9]$ (where $a$
and $m$ are the spin and mass parameters of the black hole, respectively),
that includes cases shown to be of interest $a/m\in[0.6,0.9]$ in
simulations of coalescences of neutron stars \cite{BR2017,Kastaun2013}.

\section{Equations}

We assume a \emph{stationary} metric of the form 
\begin{align}
ds^{2} & =-\alpha^{2}{dt}^{2}+r^{2}\sin^{2}\theta\psi^{4}\left(d\varphi+\beta dt\right)^{2}\nonumber \\
 & \qquad+\psi^{4}e^{2q}\left(dr^{2}+r^{2}d\theta^{2}\right). \label{metric}
\end{align}
Here $t$ is the time coordinate, and $r$, $\theta$, $\varphi$
are spherical coordinates. The gravitational constant $G=1$ and the
speed of light $c=1$. We assume axial symmetry and employ the stress-momentum
tensor 
\[
T^{\alpha\beta}=\rho hu^{\alpha}u^{\beta}+pg^{\alpha\beta},
\]
where $\rho$ is the baryonic rest-mass density, $h$ is the specific
enthalpy, and $p$ is the pressure. Metric functions $\alpha(r,\theta)$,
$\psi(r,\theta)$, $q(r,\theta)$ and $\beta(r,\theta)$ in \eqref{metric}
depend on $r$ and $\theta$ only.

The following method can be applied to any barotropic equation of
state but we will deal with polytropes $p(\rho)=K\rho^{\gamma}$.
Then one has the specific enthalpy 
\[
h(\rho)=1+\frac{\gamma p}{(\gamma-1)\rho}.
\]
The 4-velocity $(u^{\alpha})=(u^{t},0,0,u^{\varphi})$ is normalized,
$g_{\alpha\beta}u^{\alpha}u^{\beta}=-1$. The coordinate angular velocity
reads 
\begin{equation}
\Omega=\frac{u^{\varphi}}{u^{t}}.\label{Omega_def}
\end{equation}

It is well known that general-relativistic Euler equations are solvable
under the integrability condition that $j\equiv j(\Omega)$ depends
only on the angular velocity \cite{Butterworth_Ipser,Bardeen_1970}.
Within the fluid region, the Euler equations $\nabla_{\mu}T^{\mu\nu}=0$
can be integrated, yielding 
\[
\int hu_{\varphi}d\Omega+\frac{h}{u^{t}}=C_{1}
\]
or 
\begin{equation}
\int u^{t}u_{\varphi}d\Omega+\ln\left(\frac{h}{u^{t}}\right)=C_{2}.\label{uf}
\end{equation}
We choose the latter form, and define the angular momentum per unit
inertial mass $\rho h$ \cite{FM} 
\begin{equation}
j\equiv u_{\varphi}u^{t}.\label{j_def}
\end{equation}
In \cite{MM} we have had the rotation law 
\begin{equation}
j(\Omega)\equiv\frac{\tilde{w}^{1-\delta}\Omega^{\delta}}{1-\kappa\tilde{w}^{1-\delta}\Omega^{1+\delta}+\Psi};\label{prime}
\end{equation}
here $\Psi$ is of the order of the binding energy per unit baryonic
mass. This law followed from an ``educated guess-work,'' in which
three elements played equally important roles: the rotation law should
have the right (monomial) Newtonian limit, it should yield the right
1PN correction to the angular velocity, and the Bernoulli equation
should have a unique form. The last demand was ensured by imposing
the condition that a massless disk of dust exactly satisfies the Einstein-Bernoulli
equations. The full reasoning is described in \cite{MM}.

Simple rescaling 
\[
w^{1-\delta}=\frac{\tilde{w}^{1-\delta}}{1+\Psi}
\]
transforms (\ref{prime}) into 
\[
j(\Omega)\equiv\frac{w^{1-\delta}\Omega^{\delta}}{1-\kappa w^{1-\delta}\Omega^{1+\delta}}=\left(-\kappa\,\Omega+w^{\delta-1}\Omega^{-\delta}\right)^{-1},
\]
where $w$, $\delta$, and $\kappa=(1-3\delta)/(1+\delta)$ are parameters.
Thus the rotation law reads 
\begin{equation}
j(\Omega)\equiv\left(-\kappa\,\Omega+w^{\delta-1}\Omega^{-\delta}\right)^{-1},\label{momentum}
\end{equation}
where $w$, $\delta$, and $\kappa$ are parameters. The rotation curves
$\Omega(r,z)$ ought to be recovered from the Eq. (\ref{j_def}),
which can be written as 
\begin{equation}
j(\Omega)=\frac{V^{2}}{\left(\Omega+\beta\right)\left(1-V^{2}\right)},\label{rotation_law}
\end{equation}
where the square of the linear velocity reads 
\[
V^{2}=r^{2}\sin^{2}\theta\left(\Omega+\beta\right)^{2}\frac{\psi^{4}}{\alpha^{2}},
\]
and $j(\Omega)$ is defined by \eqref{momentum}. The integro-algebraic
Bernoulli equation (\ref{uf}) is given by a simple algebraic form
\cite{MM} 
\begin{eqnarray}
h\alpha\sqrt{1-V^{2}}\left(1-\kappa w^{1-\delta}\Omega^{1+\delta}\right)^{-\frac{1}{\left(1+\delta\right)\kappa}}=C.\label{algebraic_Bernoulli}
\end{eqnarray}

The central black hole is surrounded by a minimal two-surface $S_{\mathrm{BH}}$
located at $r=r_{\mathrm{s}}$ in the puncture method \cite{BrandtSeidel1995},
on a fixed hypersurface of constant time. Its area defines the irreducible
mass $M_{\mathrm{irr}}=\sqrt{\frac{A_{\mathrm{H}}}{16\pi}}$ and its
angular momentum $J_{\mathrm{H}}$ follows from the Komar expression
\begin{equation}
J_{\mathrm{H}}=\frac{1}{4}\int_{0}^{\pi/2}\frac{r^{4}\psi^{6}}{\alpha}\partial_{r}\beta\sin^{3}\theta d\theta.
\end{equation}
The angular momentum is prescribed rigidly on the event horizon $S_{\mathrm{BH}}$\textemdash it
is given by data taken from the Kerr solution (with two parameters,
mass $m$ and the spin parameter $a=J_{\mathrm{H}}/m$) and it is
independent of the content of mass in a torus. The mass of the black
hole is then defined as 
\begin{equation}
M_{\mathrm{BH}}=M_{\mathrm{irr}}\sqrt{1+\frac{J_{\mathrm{H}}^{2}}{4M_{\mathrm{irr}}^{4}}}.
\end{equation}
Another possible choice of the black hole mass is in terms of the
circumferential radius $r_{\mathrm{C}}\equiv r\sin\theta\psi^{2}$
of $S_{\mathrm{BH}}$ at the symmetry plane $\theta=\pi/2$: $M_{\mathrm{C}}=r_{\mathrm{C}}/2$.
We observed that in our numerical calculations $M_{\mathrm{C}}$ and
$M_{\mathrm{BH}}$ differ by significantly less than 1\%; this is
consistent with findings of Shibata \cite{MSH}. The asymptotic mass
$M_{\mathrm{ADM}}$ is defined as in \cite{MSH}. We decided to define
the mass of tori by $m_{\mathrm{T}}\equiv M_{\mathrm{ADM}}-M_{\mathrm{BH}}$.
The characteristic feature of this construction is that the metric
of the whole spacetime coincides with the Kerr metric in the limit
of $m_{\mathrm{T}}\to0$.

Let $K_{ij}$ denote the extrinsic curvature of the $t=\mathrm{const}$
hypersurface. Define the conformal extrinsic curvature $\hat{K}_{ij}$
as $\hat{K}_{ij}=\psi^{2}K_{ij}$. The only nonzero component $\beta$
of the shift vector is split as $\beta=\beta_{\mathrm{K}}+\beta_{\mathrm{T}}$,
where $\beta_{\mathrm{K}}$ and $\beta_{\mathrm{T}}$ are determined
as follows. The nonvanishing components of $\hat{K}_{ij}$ can be
written in the form 
\[
\hat{K}_{r\varphi}=\frac{H_{\mathrm{E}}\sin^{2}\theta}{r^{2}}+\frac{\psi^{6}}{2\alpha}r^{2}\sin^{2}\theta\partial_{r}\beta_{\mathrm{T}},
\]
\[
\hat{K}_{\theta\varphi}=\frac{H_{\mathrm{F}}\sin\theta}{r}+\frac{\psi^{6}}{2\alpha}r^{2}\sin^{2}\theta\partial_{\theta}\beta_{\mathrm{T}}.
\]
As in \cite{MSH}, we choose the functions $H_{\mathrm{E}}$ and $H_{\mathrm{F}}$
to be expressed by the formulas obtained for the Kerr metric of mass
$m$ and the spin parameter $a$, written in the form (\ref{metric}).
In explicit terms they read \cite{BrandtSeidel1995} 
\[
H_{\mathrm{E}}=\frac{ma\left[(r_{\mathrm{K}}^{2}-a^{2})\Sigma_{\mathrm{K}}+2r_{\mathrm{K}}^{2}(r_{\mathrm{K}}^{2}+a^{2})\right]}{\Sigma_{\mathrm{K}}^{2}},
\]
\[
H_{\mathrm{F}}=-\frac{2ma^{3}r_{\mathrm{K}}\sqrt{r_{\mathrm{K}}^{2}-2mr_{\mathrm{K}}+a^{2}}\cos\theta\sin^{2}\theta}{\Sigma_{\mathrm{K}}^{2}},
\]
where 
\[
r_{\mathrm{K}}=r\left(1+\frac{m}{r}+\frac{m^{2}-a^{2}}{4r^{2}}\right),
\]
and 
\[
\Sigma_{\mathrm{K}}=r_{\mathrm{K}}^{2}+a^{2}\cos^{2}\theta.
\]
It follows that for the Kerr metric, one has 
\[
\hat{K}_{r\varphi}=\frac{H_{\mathrm{E}}\sin^{2}\theta}{r^{2}}
\]
and 
\[
\hat{K}_{\theta\varphi}=\frac{H_{\mathrm{F}}\sin\theta}{r}.
\]
In the presence of the torus $\beta_{\mathrm{K}}$ has to be computed
from the relation 
\begin{equation}
\frac{\partial\beta_{\mathrm{K}}}{\partial r}=\frac{2H_{\mathrm{E}}\alpha}{r^{4}\psi^{6}}.\label{eqbetak}
\end{equation}

In what follows we proceed with the puncture method as implemented
in \cite{MSH}. Define $\Phi=\alpha\psi$ and assume the puncture
at $r=0$. Define $r_{\mathrm{s}}=\frac{1}{2}\sqrt{m^{2}-a^{2}}$,
and 
\[
\psi=\left(1+\frac{r_{\mathrm{s}}}{r}\right)e^{\phi},\quad\Phi=\left(1-\frac{r_{\mathrm{s}}}{r}\right)e^{-\phi}B.
\]
The surface $r=r_{\mathrm{s}}$ is an apparent horizon.

Einstein equations can be written as 
\begin{widetext}
\begin{subequations} \label{main_sys} 
\begin{eqnarray}
\left[\partial_{rr}+\frac{1}{r}\partial_{r}+\frac{1}{r^{2}}\partial_{\theta\theta}\right]q & = & S_{q},\label{47}\\
\left[\partial_{rr}+\frac{2r}{r^{2}-r_{\mathrm{s}}^{2}}\partial_{r}+\frac{1}{r^{2}}\partial_{\theta\theta}+\frac{\cot{\theta}}{r^{2}}\partial_{\theta}\right]\phi & = & S_{\phi},\label{44}\\
\left[\partial_{rr}+\frac{3r^{2}+r_{\mathrm{s}}^{2}}{r(r^{2}-r_{\mathrm{s}}^{2})}\partial_{r}+\frac{1}{r^{2}}\partial_{\theta\theta}+\frac{2\cot{\theta}}{r^{2}}\partial_{\theta}\right]B & = & S_{B},\label{45}\\
\left[\partial_{rr}+\frac{4r^{2}-8r_{\mathrm{s}}r+2r_{\mathrm{s}}^{2}}{r(r^{2}-r_{\mathrm{s}}^{2})}\partial_{r}+\frac{1}{r^{2}}\partial_{\theta\theta}+\frac{3\cot{\theta}}{r^{2}}\partial_{\theta}\right]\beta_{\mathrm{T}} & = & S_{\beta_{\mathrm{T}}},\label{46}
\end{eqnarray}
\end{subequations} where source terms $S_{\phi},S_{B},S_{\beta_{\mathrm{T}}},S_{q}$
are \begin{subequations} 
\begin{flalign}
 & S_{q}=-8\pi e^{2q}\left(\psi^{4}p-\frac{\rho hu_{\varphi}^{2}}{r^{2}\sin^{2}\theta}\right)+\frac{3A^{2}}{\psi^{8}}+2\left[\frac{r-r_{\mathrm{s}}}{r(r+r_{\mathrm{s}})}\partial_{r}+\frac{\cot\theta}{r^{2}}\partial_{\theta}\right]b+\left[\frac{8r_{\mathrm{s}}}{r^{2}-r_{\mathrm{s}}^{2}}+4\partial_{r}(b-\phi)\right]\partial_{r}\phi\\
 & +\frac{4}{r^{2}}\partial_{\theta}\phi\partial_{\theta}(b-\phi),\nonumber \\
 & S_{\phi}=-2\pi e^{2q}\psi^{4}\left[\rho_{\mathrm{H}}-p+\frac{\rho hu_{\varphi}^{2}}{\psi^{4}r^{2}\sin^{2}\theta}\right]-\frac{A^{2}}{\psi^{8}}-\partial_{r}\phi\partial_{r}b-\frac{1}{r^{2}}\partial_{\theta}\phi\partial_{\theta}b-\frac{1}{2}\left[\frac{r-r_{\mathrm{s}}}{r(r+r_{\mathrm{s}})}\partial_{r}b+\frac{\cot\theta}{r^{2}}\partial_{\theta}b\right],\\
 & S_{B}=16\pi Be^{2q}\psi^{4}p,\\
 & S_{\beta_{\mathrm{T}}}=\frac{16\pi\alpha e^{2q}J}{r^{2}\sin^{2}\theta}-8\partial_{r}\phi\partial_{r}\beta_{\mathrm{T}}+\partial_{r}b\partial_{r}\beta_{\mathrm{T}}-8\frac{\partial_{\theta}\phi\partial_{\theta}\beta_{\mathrm{T}}}{r^{2}}+\frac{\partial_{\theta}b\partial_{\theta}\beta_{\mathrm{T}}}{r^{2}}
\end{flalign}
\end{subequations}
\end{widetext}

and 
\[
A^{2}=\frac{\hat{K}_{r\varphi}^{2}}{r^{2}\sin^{2}\theta}+\frac{\hat{K}_{\theta\varphi}^{2}}{r^{4}\sin^{2}\theta},
\]
\[
\rho_{\mathrm{H}}=\rho h(\alpha u^{t})^{2}-p,
\]
\[
J=\rho h\alpha u^{t}u_{\varphi},
\]
\[
B=e^{b}.
\]

In the rest of the paper we always assume that $\Omega>0$. Corotating
disks have $a>0$, while counterrotating disks have negative spins:
$a<0$.

\section{Description of numerics}

The core of our numerical method is close to that described in \cite{MSH}.
Main modifications introduced in our scheme are due to the implementation
of a different rotation law which requires solving Eq.\ (\ref{rotation_law}).
The Kerr solution plays an essential role in the presented model.
It appears formally in the limit $m_{\mathrm{T}}\to0$. It is important
to stress that the assumed rotation law (\ref{momentum}) actually
admits solutions with an arbitrarily small mass. The latter property
is by no means obvious; it is for instance absent for the rigid rotation
\cite{petroff}. Moreover, the Kerr metric is the only analytic solution
that is available for testing the correctness of our numerical code.
Consequently, we implemented some boundary conditions in the way that
yields the Kerr solution with the best accuracy in the absence of
the torus. Finally, there is a technical difference in treating Eq.
(\ref{46}).

The solutions are found iteratively, using a fixed-point method. Equations
(\ref{main_sys}) are solved with respect to functions $\phi$, $B$,
$\beta_{\mathrm{T}}$, and $q$, respectively. In each iteration Eqs.\ (\ref{eqbetak})
is integrated yielding $\beta_{\mathrm{K}}$. The specific enthalpy
is computed from the Bernoulli Eq.\ (\ref{algebraic_Bernoulli}).
Our prescription of the rotation law in the form $j=j(\Omega)$ requires
also solving Eq.\ (\ref{rotation_law}) with respect to $\Omega$.

The baryonic density $\rho$ is computed from the specific enthalpy
as 
\[
\rho=\left[\frac{\gamma-1}{K\gamma}(h-1)\right]^{\frac{1}{\gamma-1}}.
\]
In each iteration we set the polytropic constant $K$ so that the
maximum of the density is equal to an \textit{a priori} specified
value $\rho_{\mathrm{max}}$; this is essentially the old idea of
Hachisu \cite{Hachisu}, to which we added a new element. We observed
that the convergence of the fixed point method significantly improves,
i.e., solutions can be obtained for a much wider range of parameters,
if the maximum of the enthalpy is sought within the whole volume of
the disk, and not only at the equatorial plane. On the other hand,
in all cases investigated so far, the final solution is characterized
by the maximum of the specific enthalpy occurring at the plane $\theta=\pi/2$.
The constants $w$ and $C$ appearing in the rotation law and Eq.\ (\ref{algebraic_Bernoulli})
are hard to control, and they are not specified \textit{a priori}.
Instead, we assume the values of the inner and outer equatorial radii
of the disk ($r_{1}$ and $r_{2}$, respectively). The values of $w$
and $C$ are obtained in each iteration from Eqs.\ (\ref{rotation_law})
and (\ref{algebraic_Bernoulli}) taken at points with the coordinates
$(r,\theta)$ equal to $(r_{1},\pi/2)$ and $(r_{2},\pi/2)$. They
constitute a set of four algebraic equations for four unknowns: $w$,
$C$, and the values of $\Omega$ at $(r_{1},\pi/2)$ and $(r_{2},\pi/2)$.
Fortunately, simple substitutions are possible; they lead effectively
to an algebraic set of two equations, which we solve using a standard
Newton-Raphson method.

Our numerical grid spans the region $r_{\mathrm{s}}\le r\le r_{\infty}$,
$0\le\theta\le\pi/2$, where $r_{\infty}$ is finite, but large: $r_{\infty}\gg r_{2}$.
We assume equatorial symmetry. This implies that $\partial_{\theta}\phi=\partial_{\theta}B=\partial_{\theta}\beta_{\mathrm{T}}=\partial_{\theta}q=0$
at $\theta=\pi/2$. On the axis we assume regularity conditions, that
is $\partial_{\theta}\phi=\partial_{\theta}B=\partial_{\theta}\beta_{\mathrm{T}}=0$
at $\theta=0$. We also put $q=0$ at $\theta=0$, which is required
by the local flatness of the metric. The boundary conditions at $r=r_{\mathrm{s}}$
follow from the puncture construction. We require that $\partial_{r}\phi=\partial_{r}B=\partial_{r}\beta_{\mathrm{T}}=\partial_{r}q=0$
at $r=r_{\mathrm{s}}$. It is observed in \cite{MSH} that Eq.\ (\ref{46})
admits a more stringent condition at $r=r_{\mathrm{s}}$. We follow
\cite{MSH} and require $\beta_{\mathrm{T}}=O[(r-r_{\mathrm{s}})^{4}]$,
or equivalently $\partial_{r}\beta_{\mathrm{T}}=\partial_{rr}\beta_{\mathrm{T}}=\partial_{rrr}\beta_{\mathrm{T}}=0$
at $r=r_{\mathrm{s}}$. To some extent this choice is arbitrary and
it is connected with the freedom of defining the splitting of the
shift vector $\beta=\beta_{\mathrm{T}}+\beta_{\mathrm{K}}$, but it
has consequences in the definition and properties of the angular momentum
of the black hole.

The conditions at the outer boundary follow from the multipole expansion
and the conditions of asymptotic flatness. We have 
\begin{eqnarray}
\phi & \sim & \frac{M_{1}}{2r}, \ \ \ \ \ B  \sim 1-\frac{B_{1}}{r^{2}}, \nonumber \\
\beta_{\mathrm{T}} & \sim & -\frac{2J_{1}}{r^{3}},  \ \ \ \ \ q  \sim  \frac{q_{1}\sin^{2}\theta}{r^{2}},\label{boundinf}
\end{eqnarray}
as $r\to\infty$, where the constants $M_{1}$, $B_{1}$, $J_{1}$ and $q_{1}$ satisfy 
\begin{equation}
M_{1}=-2\int_{r_{\mathrm{s}}}^{\infty}(r^{2}-r_{\mathrm{s}}^{2})dr\int_{0}^{\pi/2}\sin\theta d\theta S_{\phi},\label{m1}
\end{equation}
\begin{equation}
B_{1}=\frac{2}{\pi}\int_{r_{\mathrm{s}}}^{\infty}dr\frac{(r^{2}-r_{\mathrm{s}}^{2})^{2}}{r}\int_{0}^{\pi/2}d\theta\sin^{2}\theta S_{B},\label{b1}
\end{equation}
\begin{equation}
J_{1}=4\pi\int_{r_{\mathrm{s}}}^{\infty}r^{2}dr\int_{0}^{\pi/2}\sin\theta d\theta\rho\alpha u^{t}\psi^{6}e^{2q}hu_{\varphi},\label{j1}
\end{equation}
\begin{eqnarray}
q_{1} & = & \frac{2}{\pi}\int_{r_{\mathrm{s}}}^{\infty}drr^{3}\int_{0}^{\pi/2}d\theta\cos(2\theta)S_{q}\nonumber \\
 &  & -\frac{4}{\pi}r_{\mathrm{s}}^{2}\int_{0}^{\pi/2}d\theta\cos(2\theta)q(r_{\mathrm{s}},\theta).\label{q1}
\end{eqnarray}
Note a misprint in the boundary condition for $B$ in \cite{MSH}.

In our implementation the grid points can be distributed almost freely,
both in radial and angular directions, but we require existence of
grid points at the boundaries, i.e., for $r=r_{\mathrm{s}}$, $\theta=0$,
$\theta=\pi/2$ and $r=r_{\infty}$. In practice we distribute the
grid points in the radial direction according to 
\begin{equation}
r_{i}=r_{\mathrm{s}}+\frac{f^{i-1}-1}{f-1}\Delta r,\quad i=1,2,\dots,N_{r}\label{r_grid}
\end{equation}
where $f$ and $\Delta r$ are constants. This choice is similar to
the one used in \cite{MSH}. In the angular direction we either set
an equidistant grid 
\begin{equation}
\theta_{j}=(j-1)\Delta\theta,\quad j=1,\dots,N_{\theta},\label{eqtheta}
\end{equation}
where $\Delta\theta=\pi/[2(N_{\theta}-1)]$, or a grid with almost
equally spaced values of $\mu=\cos\theta$, i.e., 
\begin{equation}
\theta_{j}=\begin{cases}
0, & j=1,\\
\arccos\left[1+\left(\frac{3}{2}-j\right)\Delta\mu\right], & j=2,\dots,N_{\theta}-1,\\
\frac{\pi}{2}, & j=N_{\theta},
\end{cases}\label{eqcos}
\end{equation}
where $\Delta\mu=1/(N_{\theta}-2)$.

Equations (\ref{main_sys}) are discretized using second order (3-point)
difference formulas. The implementation of the boundary conditions
is more subtle. In each iteration we set the values of $\phi$, $B$,
$\beta_{\mathrm{T}}$ and $q$ at the boundary grid nodes, assuring
that appropriate boundary conditions are satisfied. At the outer boundary,
i.e., for $r=r_{\infty}$, we set the values of $\phi$, $B$, $\beta_{\mathrm{T}}$
and $q$ as given by Eqs.\ (\ref{boundinf}) and (\ref{m1})\textendash (\ref{q1}).
At the axis ($\theta=0$), equator ($\theta=\pi/2$), and the horizon
($r=r_{\mathrm{s}}$), the boundary conditions generally amounts to
a requirement that an appropriate first derivative vanishes. We express
this derivative using 3-, 4- or 6-point formulas, and set the boundary
values appropriately. These boundary values are treated as fixed in
the process of solving Eq.\ (\ref{main_sys}) for $\phi$, $B$,
$\beta_{\mathrm{T}}$, and $q$ in the bulk of the grid. In other words,
in each iteration, Eqs.\ (\ref{main_sys}) are solved as if they
constituted a Dirichlet boundary problem. This approach seems to be
essential for the convergence properties of the iterative fixed-point
scheme. The only exception from this rule is our treatment of the
boundary condition for $\beta_{\mathrm{T}}$ at the horizon. In the
latter case the equations expressing the conditions $\partial_{r}\beta_{\mathrm{T}}=\partial_{rr}\beta_{\mathrm{T}}=\partial_{rrr}\beta_{\mathrm{T}}=0$
at $r=r_{\mathrm{s}}$ are solved together with Eq.\ (\ref{46})
using 6-point finite difference formulae. Other boundary conditions
are implemented using 3-point formulas. A 4-point finite difference
formula is used only for the condition $\partial_{r}\phi=0$ at $r=r_{\mathrm{s}}$.
\textit{We would like to stress that this choice yielded the best
accuracy in the test consisting in recovering the Kerr solution in
the absence of the torus.} On the other hand, it is not optimal with
respect to the computation of the $R_{\mathrm{con}}$ parameter, which
we describe in Sec.\ IV A.

All integrals (\ref{m1})\textendash (\ref{q1}) are computed using
standard trapezoid formulae in each iteration. Trapezoid formulas
are also used in the integration of Eq.\ (\ref{eqbetak}) for $\beta_{\mathrm{K}}$.
Following \cite{MSH}, we assume that $\beta_{\mathrm{K}}=0$ at $r=r_{\infty}$,
which is understood as an approximation of the boundary condition
$\beta_{\mathrm{K}}\to0$ for $r\to\infty$.

Most of the computing time is spent on solving the discretized versions
of Eqs.\ (\ref{main_sys}). We exploit the banded matrix structure
of these four equations, and solve them performing standard $LU$
matrix decomposition \cite{banachiewicz} using LAPACK \cite{lapa}.
A technical change as compared to the formulation of \cite{MSH} is
that we have moved terms depending on the derivatives of $\phi$ and
$b$ to the right-hand side of Eq.\ (\ref{46}). As a result the
coefficients of the operators on the left-hand sides of Eq.\ (\ref{main_sys})
do not change during iterations, and it is possible to perform the
decomposition of the matrices once only. Equation (\ref{rotation_law})
is solved in each iteration with respect to $\Omega$, using a Newton-Raphson
method. We search for the solution for $\Omega$ in a bounded region
containing the torus.

Choosing appropriate initial data is an inherent problem of all iterative
methods for obtaining relativistic figures of equilibrium. Rotation
law (\ref{momentum}) admits starting from the Kerr data, i.e., there
exist an acceptable solution of Eq.\ (\ref{algebraic_Bernoulli})
for the Kerr metric, except for the case with $a=0$, and $\kappa=3$.
We obtain the solutions of the latter case by first obtaining a solution
for $\kappa=3$ and a small, but nonzero $a$. This solution is then
used as a starting point to obtain the final solution for $\kappa=3$
and $a=0$.

In addition to the parameters of the rotation law {[}$\delta$ and
$\kappa$ in case of the rotation law (\ref{momentum}){]}, the solution
is specified by the following six parameters: the mass $m$ and the
spin $a$ parameters of the black hole; the inner and outer coordinate
radii of the disk at the equatorial plane: $r_{1}$ and $r_{2}$,
respectively; the maximum of the baryonic density within the disk
$\rho_{\mathrm{max}}$; the polytropic exponent $\gamma$. We assume
in what follows $m=1$ and $\gamma=4/3$. All length related quantities
($r_{1}$, $r_{2}$ and circumferential radii) shall be expressed
in terms of $m$.

We should comment here on terminology. As we have pointed above, in
the construction of Shibata \cite{MSH} the angular momentum of the
black hole is defined \textit{a priori} in terms of parameters $a$
and $m$: the ``dressed'' angular momentum is the same as the naked
one, $J_{\mathrm{BH}}=am$. On the other hand the ``dressed'' mass
of a black hole $M_{\mathrm{BH}}$ is different from the mass parameter
$m$. Thus $J_{\mathrm{BH}}/M_{\mathrm{BH}}\ne a$; the notion of
the spin parameter $a$ would become ambiguous. In order to avoid
misunderstanding, when we use the term \textit{spin parameter $a$},
we always mean that $J_{\mathrm{BH}}=am$.

\section{Testing numerical codes}

We shall provide below a short discussion of the accuracy of our numerical
codes. We begin, in the first subsection, with elements of standard
numerical analysis. Then we shall compare our results on two metric
functions with those of \cite{MSH}.

\subsection{Numerical tests}

From the mathematical point of view, a solution for a given rotation
law {[}with one free parameter, e.g.,\ \eqref{momentum} with fixed
$\kappa${]}, is expected to be defined by ``physical'' parameters:
$m$, $a$, $r_{1}$, $r_{2}$, $\rho_{\mathrm{max}}$. This solution
might not be unique, in principle. However, numerical solutions depend
on several additional parameters. The most obvious is the grid size
$N_{r}\times N_{\theta}$ and its spacing. Our mesh is rectangular
in the radial and angular variables. The inner edge in radius is by
construction always positioned at the apparent horizon $r_{\mathrm{s}}$,
but the outer edge radius $r_{\infty}$ provides another numerical
parameter. It might influence asymptotic expansions used to define
outer boundary conditions, see Eqs.\ (\ref{boundinf})\textendash (\ref{q1}).
Noteworthy, it is directly related to the grid spacing $\Delta r$
near the horizon (\ref{r_grid}). Less obvious numerical errors come
from the surface-capturing scheme, which does not allow one to define
the free boundary with the accuracy better than the size of a cell.
In particular, in our implementation inner and outer radii $r_{1}$,
$r_{2}$ are pushed upward to nearest equatorial mesh points and the
aliasing error is inevitable. Finally, the ``fixed-point'' in practice
is never reached. The convergence is limited by a numerical noise,
albeit it is of a tiny amplitude. The forthcoming analysis will quantify
dependence of the results on variations of $N_{r},N_{\theta}$, $r_{\mathrm{\infty}}$
or $\Delta r$ and the number of iterations. Grids as small as $N_{r}\times N_{\theta}=17\times9$
and as big as $N_{r}\times N_{\theta}=8193\times257$ were used in
testing procedures. The lowest resolution yields results that are
too inaccurate but the numerical solution still exists. This demonstrates
robustness and stability of the numerical code.

In the next three subsections we put $a=0.6$, $r_{1}=8$, $r_{2}=20$,
$\rho_{\mathrm{max}}=4\times10^{-4}$ and use the rotation law (\ref{momentum})
with $\delta=-1/3$ and $\kappa=3$. We applied the geometric variant
of (\ref{r_grid}), i.e., with $\Delta r=(f-1)r_{\mathrm{s}}$; this
implies $r_{i}=r_{\mathrm{s}}f^{i-1}$ with $f=1.0113622034834149$.
That facilitates the grid doubling\footnote{By doubling we understand adding new grid points between adjacent
points of the old grid.}. The radius $r_{\infty}$ was kept constant by squaring $f$ or taking
the square root of $f$ when the grid was halved or doubled relative
to $N_{r}=511$, respectively. We used the grid equidistant in $\cos{\theta}$,
in tests concerning the angular resolution. The grid resolutions $N_{r}\times N_{\theta}$
are defined as integer powers of 2 minus 1, to keep them odd. The
machine precision $\epsilon\simeq2.22\times10^{-16}$ was shown in
Figs.~\ref{fig:Fig1}\textendash \ref{fig:Fig4} for reference. We
use the maximum norm $L^{\infty}$ to quantify errors. We measure
the difference between the values of metric functions and the enthalpy
in two subsequent iterations. One can use other matrix norms that
would give (as we have checked) values larger or smaller by a few
orders of magnitude. On the other hand error ratios and line slopes
shown in Figs. \ref{fig:Fig1}\textendash \ref{fig:Fig4} are invariant.

\subsubsection{Fixed point noise}

\begin{figure}[ht]
\includegraphics[width=0.5\textwidth]{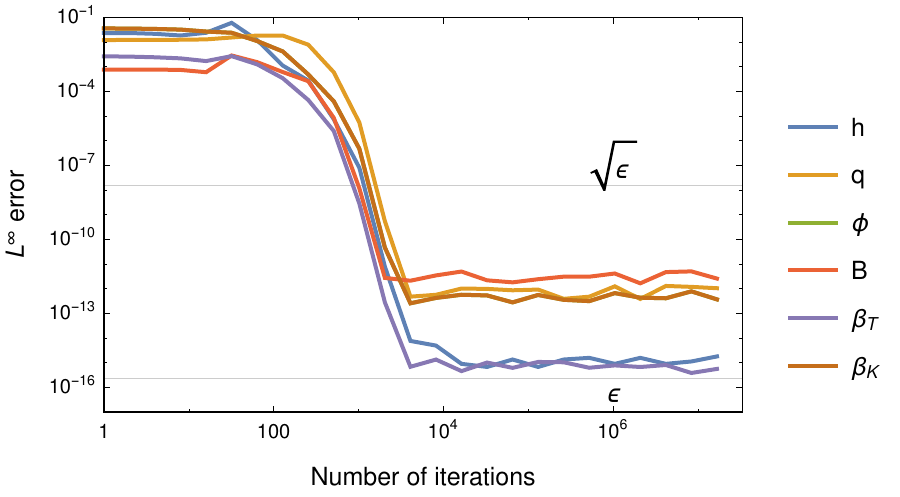} \includegraphics[width=0.5\textwidth]{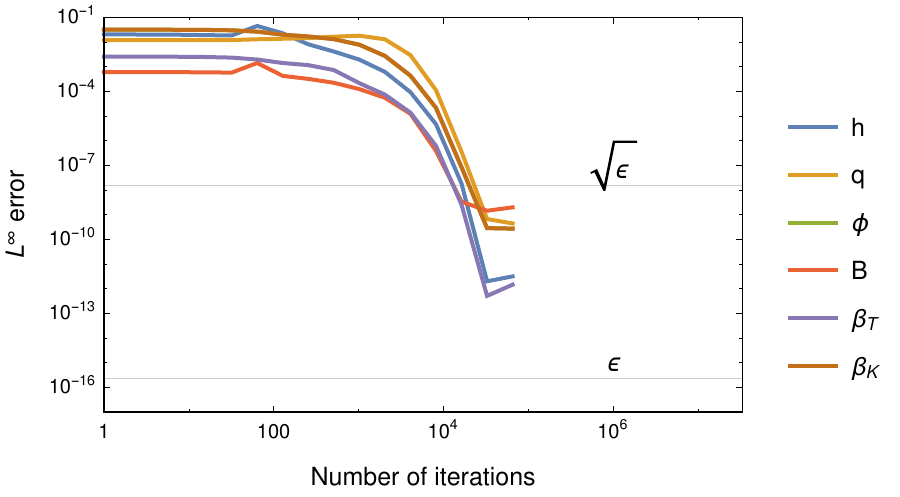}
\caption{\label{fig:Fig1} Typical behavior of the $L^{\infty}$ error measured
with respect to the preceding iteration for two extreme cases with
the grids $N_{r}\times N_{\theta}=2049\times257$ (the bottom diagram)
and $N_{r}\times N_{\theta}=65\times33$ (the upper diagram).}
\end{figure}

A typical behavior during numerical iterations (Figs.\ \ref{fig:Fig1})
consists of three stages: (1) a short (a few hundred iterations) divergent
stage with the maximum mass density occurring outside the symmetry
plane $\theta=\pi/2$; (2) a stage characterized by an exponential
convergence ($10^{3}$\textendash $10^{4}$ iterations); (3) a steady
stage with the error dominated by the numerical noise. With the increased
grid size two first stages are longer both with respect to the number
of iterations and the computational time. The amplitude of the numerical
noise is increased as well. Fortunately, in all investigated cases
the magnitude of the noise is small and of no significance. In practice,
the errors associated with other factors, which we describe below,
are always larger. Consequently, in most cases it suffices to fix
the number of iterations at a level of a few thousands.

\subsubsection{Doubling the grid: Increasing $N_{r}$ and $N_{\theta}$ with fixed
$r_{\infty}$}

\begin{figure}
\includegraphics[width=0.5\textwidth]{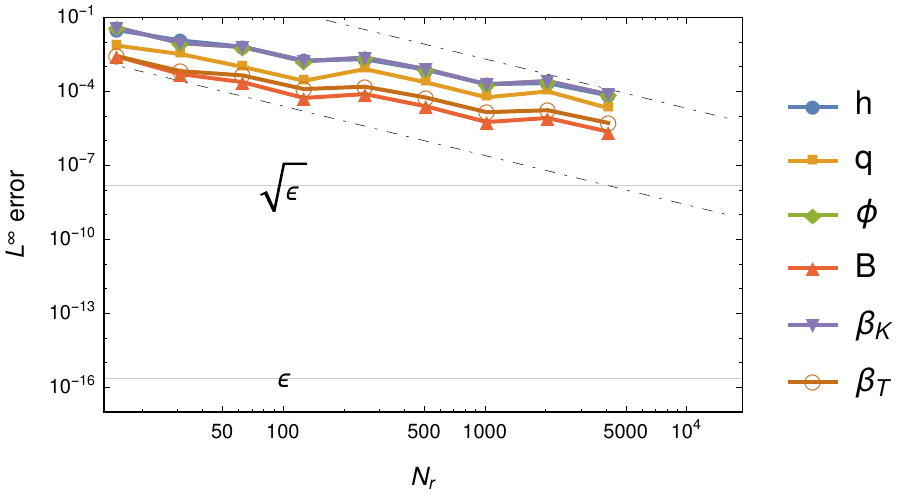} \caption{\label{fig:Fig2} Typical behavior of the $L^{\infty}$ error with
the increase of the radial grid resolution. Here $r_{\infty}$ is
constant. Lowest resolution $N_{r}\times N_{\theta}=17\times17$ was
doubled in radial direction nine times until reference model $N_{r}\times N_{\theta}=8193\times17$
(not shown) was obtained.}
\end{figure}

\begin{figure}
\includegraphics[width=0.5\textwidth]{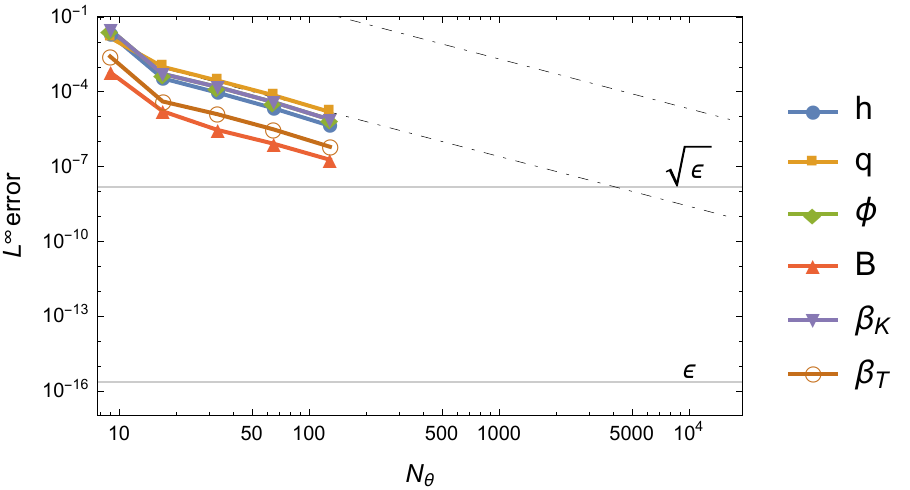} \caption{\label{fig:Fig3} The same as in Fig.~\ref{fig:Fig2}, but with the
increased angular resolution. The grid grows from $N_{r}\times N_{\theta}=1025\times9$
to $N_{r}\times N_{\theta}=1025\times257$.}
\end{figure}

By doubling the grid we simultaneously reduce two sources of errors:
(1) the discretization error; (2) the aliasing error. Typically, doubling
the radial grid gives an almost quadratic convergence (Fig.\ \ref{fig:Fig2}),
while doubling the angular grid (free of an aliasing error) yields
convergence that is exactly quadratic (cf.\ Fig.~\ref{fig:Fig3}).

\subsubsection{Extending the grid: Increasing $N_{r}$ with fixed $f$ and $\Delta r$}

\begin{figure}[b]
\includegraphics[width=0.5\textwidth]{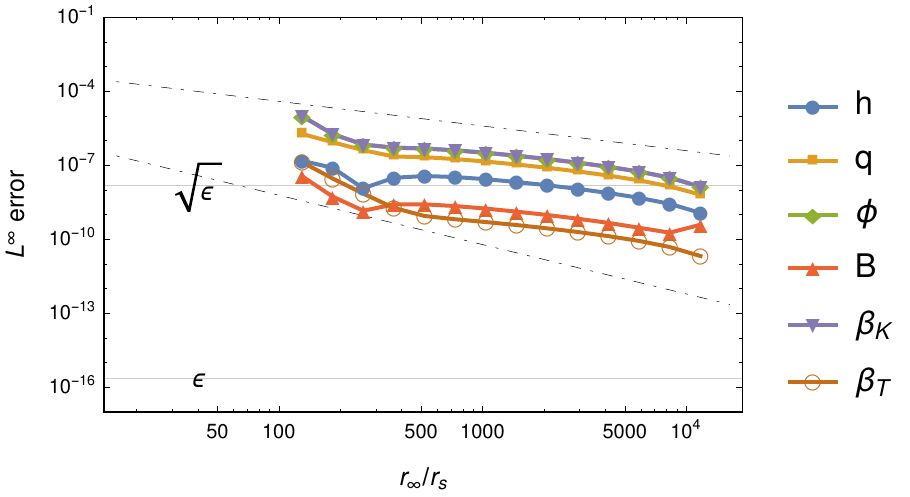} \caption{\label{fig:Fig4} Typical behavior of the $L^{\infty}$ error with
the increase of the $r_{\infty}$. Radial grid was extended from $N_{r}\times N_{\theta}=513\times65$
to $N_{r}\times N_{\theta}=961\times65$, while keeping $f$ and $\Delta r$
constant in (\ref{r_grid}). }
\end{figure}

The extension of the outer grid edge $r_{\infty}$ has a rather moderate
impact on the global error, which decreases linearly with the increase
of $r_{\infty}$ (see Fig.~\ref{fig:Fig4}).

\subsubsection{$R_{\text{con}}$ test}

\begin{figure}[b]
\includegraphics[width=0.5\textwidth]{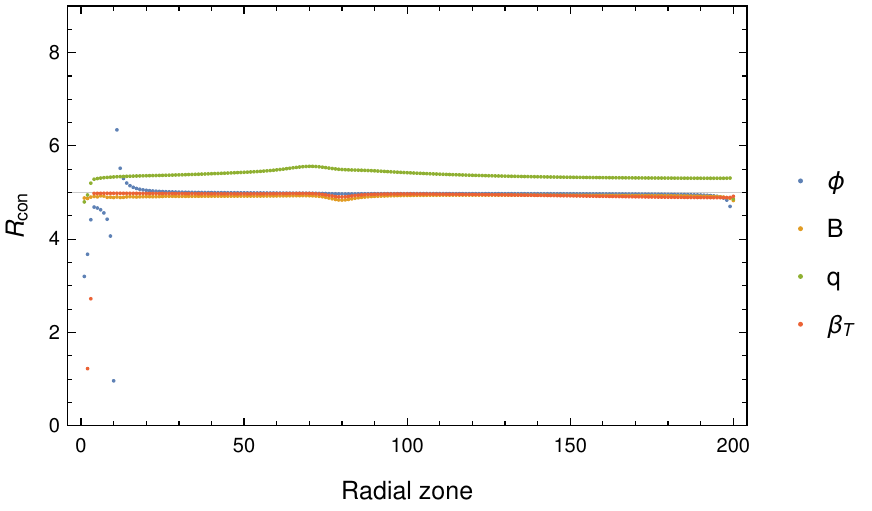} \includegraphics[width=0.5\textwidth]{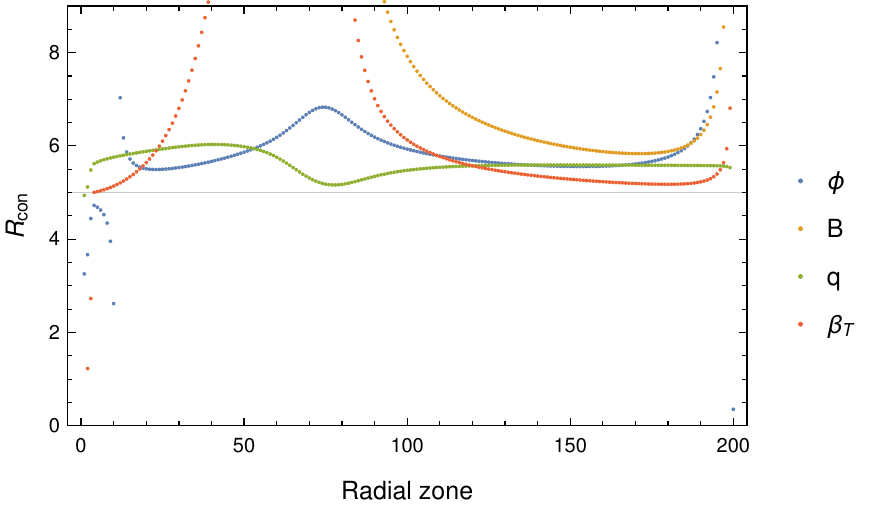}
\caption{\label{RconTEST} Results for $R_{\mathrm{con}}$ tests. In the upper
part of the figure the black hole has $a=0.9$ and the torus rotates
according to (\ref{j(O)-1}), where $\kappa=3$. In the lower part
the black hole has $a=0.9$ and the torus rotates according to (\ref{momentum1}).}
\end{figure}

In order to disentangle the discretization and aliasing errors, we
employ the test of Shibata (see the Appendix B in \cite{MSH}) for
grid-aligned tori. One can compute the value: 
\[
R_{\text{con}}=\frac{Q_{4}-Q_{1}}{Q_{2}-Q_{1}},
\]
where $Q_{1},Q_{2},Q_{4}$ denotes values obtained at some grid node
with the use of the original radial grid, a grid with every second
node removed and another grid obtained from the former by repeating
the removal operation. For the interpolation of the $n$th order one
could expect $R_{\text{con}}=2^{n}+1$, i.e., $R_{\text{con}}=5$
for the quadratic interpolation/discretization. Unfortunately, results
that have been obtained are rather inconclusive. For Shibata's test
model and for one of our models (a central black hole with the spin
parameter $a=0.9$ and the rotation law (\ref{j(O)-1}) with $a=0.9$
and $\kappa=3$) we get in fact $R_{\text{con}}\simeq5\pm0.05$, for
all metric functions with the exception of $q$ and a few grid nodes
in the vicinity of the horizon. In this test we implement boundary
conditions at the horizon for $\phi$ and $q$ using 3- and 4-point
finite difference formulas, respectively.

On the other hand, for our model with the rotation law (\ref{momentum1}),
and with the same spin parameter of the central black hole, $a=0.9$,
the situation becomes confusing. We get $R_{\text{con}}\simeq5\pm1$
at best, with a large variation for some metric functions (see Fig.\ \ref{RconTEST}).
Clarification of these issues requires further testing, but we are
not convinced that the tests based on the $R_{\text{con}}$ parameter
are necessarily useful. It is important to stress, that any value
of $R_{\text{con}}>1$ indicates the convergence. There are values
of $R_{\text{con}}\gg5$ (Fig.\ \ref{RconTEST}, lower panel) that
indicate convergence better than expected. This might be surprising,
but it is possible. For example, if one of the metric function is
well-approximated by a quadratic polynomial, the 2nd order scheme
could yield almost exact solution of the discretized system instantly.

\subsubsection{Conclusions}

Judging from the contribution to the $L^{\infty}$ norm of the error,
its main sources can be ordered as follows. The radial resolution
(including aliasing), and the angular resolution contribute ($10^{-3}$\textendash $10^{-5}$)
and ($10^{-5}$\textendash $10^{-7}$) to the budget error, respectively.
Further contributions come from the radial grid extension ($10^{-8}$\textendash $10^{-10}$)
and the fixed-point noise error ($10^{-11}$\textendash $10^{-16}$).

Apart from the convergence tests described above, we have also recovered
the Kerr metric for the vacuum case, i.e., without the torus. The
code successfully passed all these tests.

\begin{figure}[h]
\includegraphics[width=1\columnwidth]{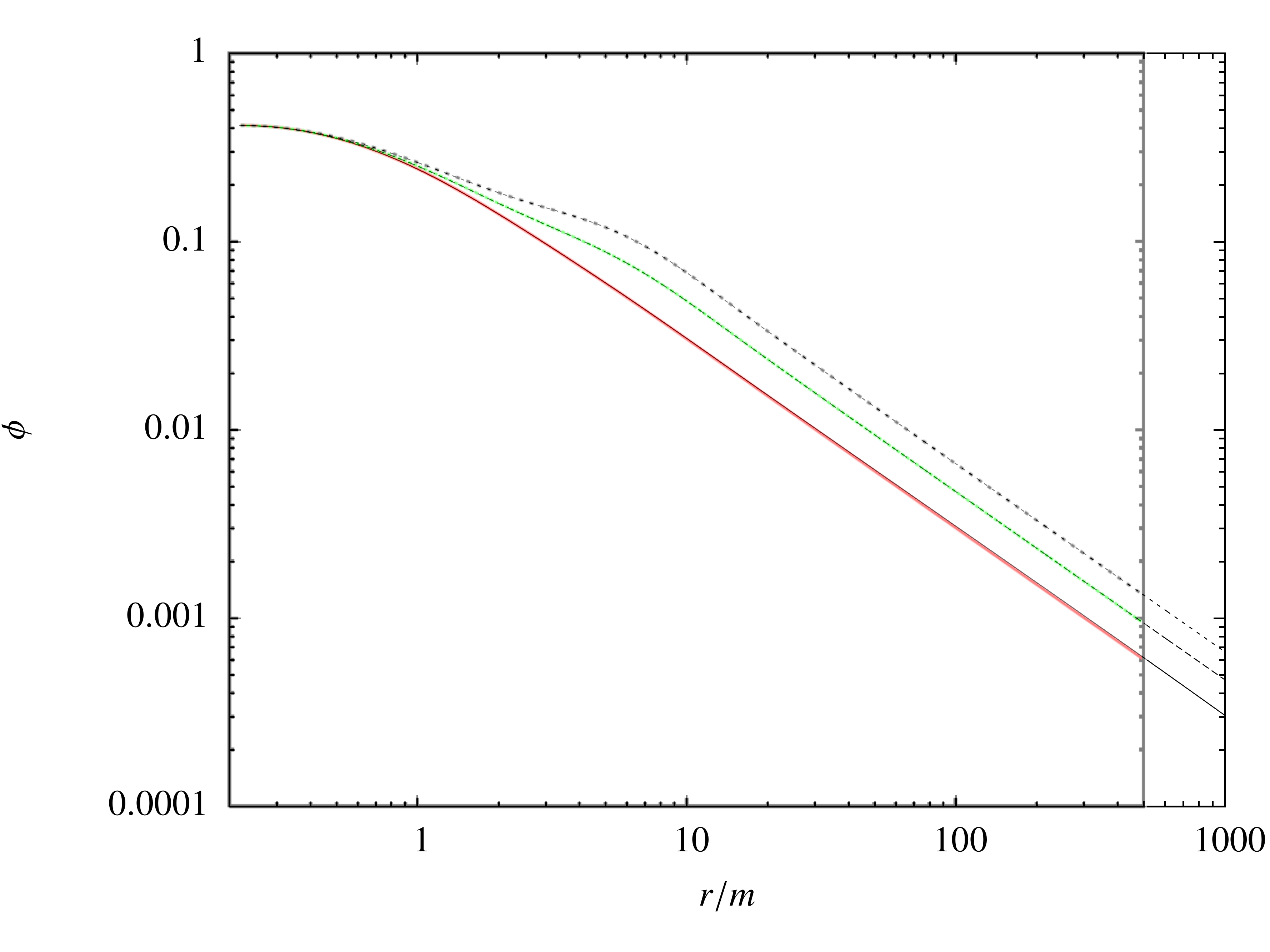} \includegraphics[width=1\columnwidth]{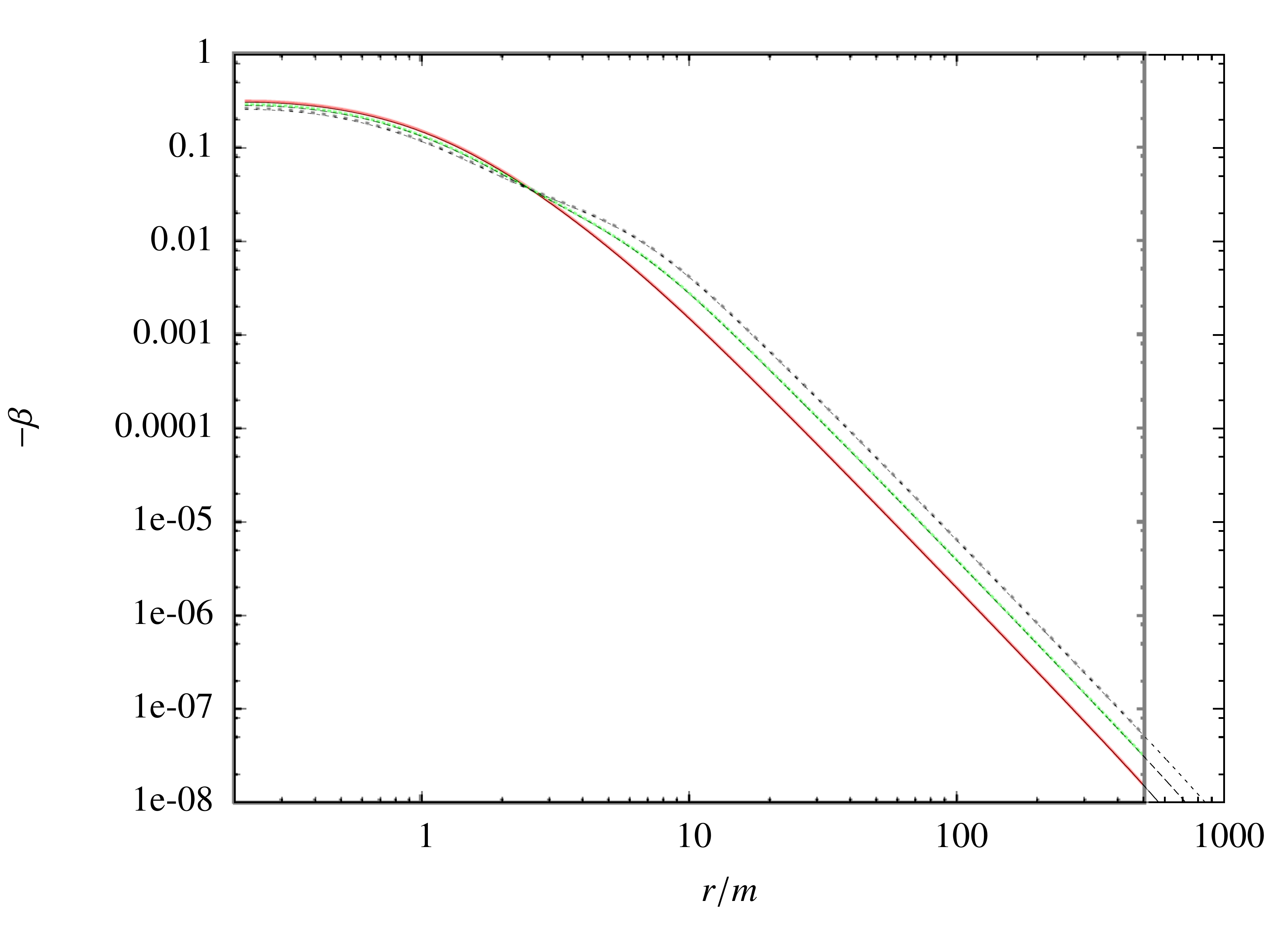}
\caption{\label{fig:por_a}Functions $\phi$ and $-\beta$ in the equatorial
plane for the configuration with the spin parameter of the black hole
$a=0.9$, $r_{1}=1.5$, $r_{2}=20.0$, for the cases $M_{\ast}/m=0.043$,
$\rho_{\mathrm{max}}=0.475\times10^{-4}$ (solid line), $M_{\ast}/m=0.400$,
$\rho_{\mathrm{max}}=3.128\times10^{-4}$ (broken line) and $M_{\ast}/m=0.800$,
$\rho_{\mathrm{max}}=4.936\times10^{-4}$ (dashed-dotted line). Inset
in the two figures (with the abscissa extending up to $r/m=500$)
are the corresponding diagrams taken from Fig. 4 of \cite{MSH}. }
\end{figure}

\subsection{Rotation with constant $\boldsymbol{hu_{\varphi}}$}

\begin{figure}[b]
\includegraphics[width=1\columnwidth]{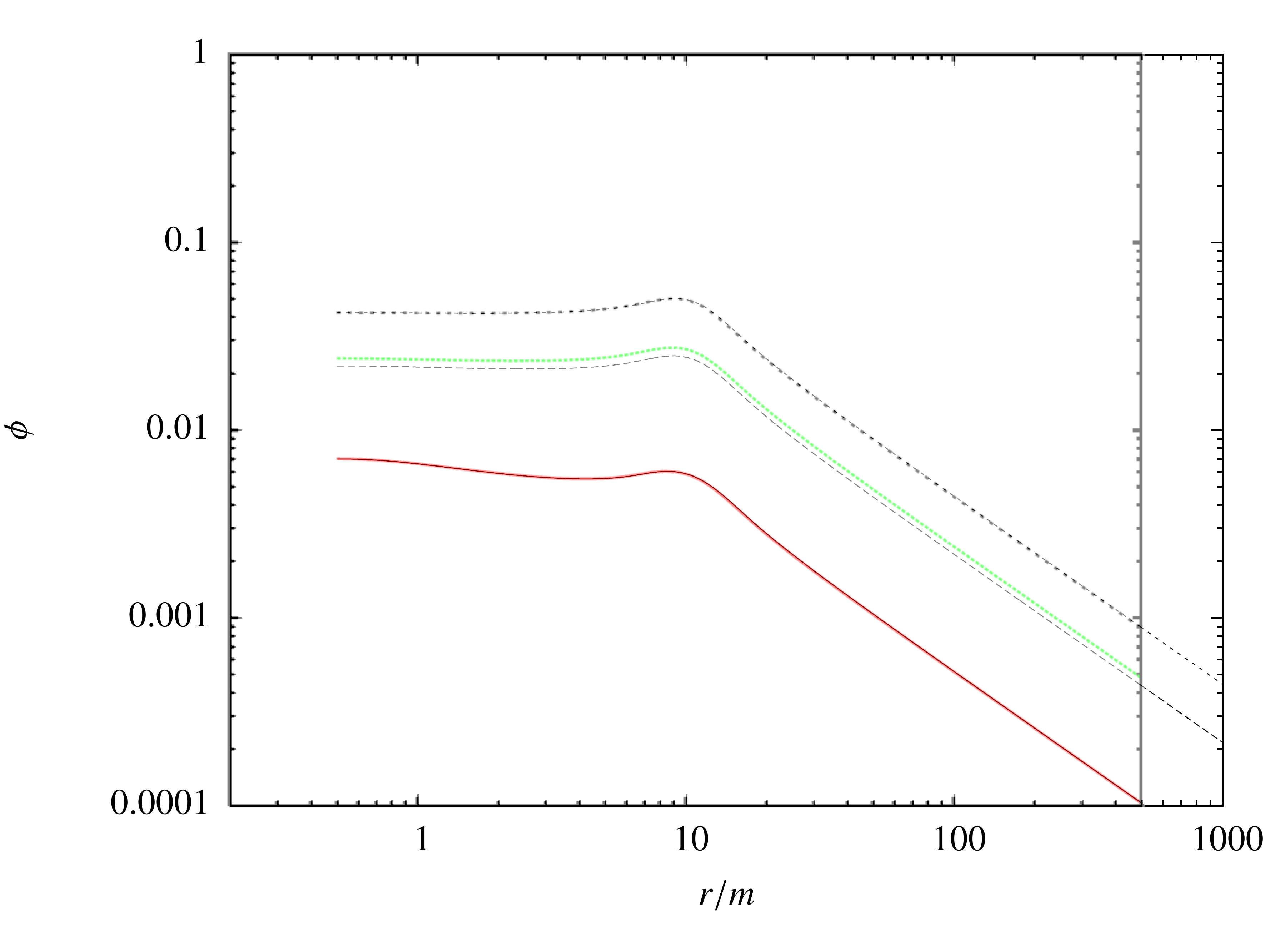} \includegraphics[width=1\columnwidth]{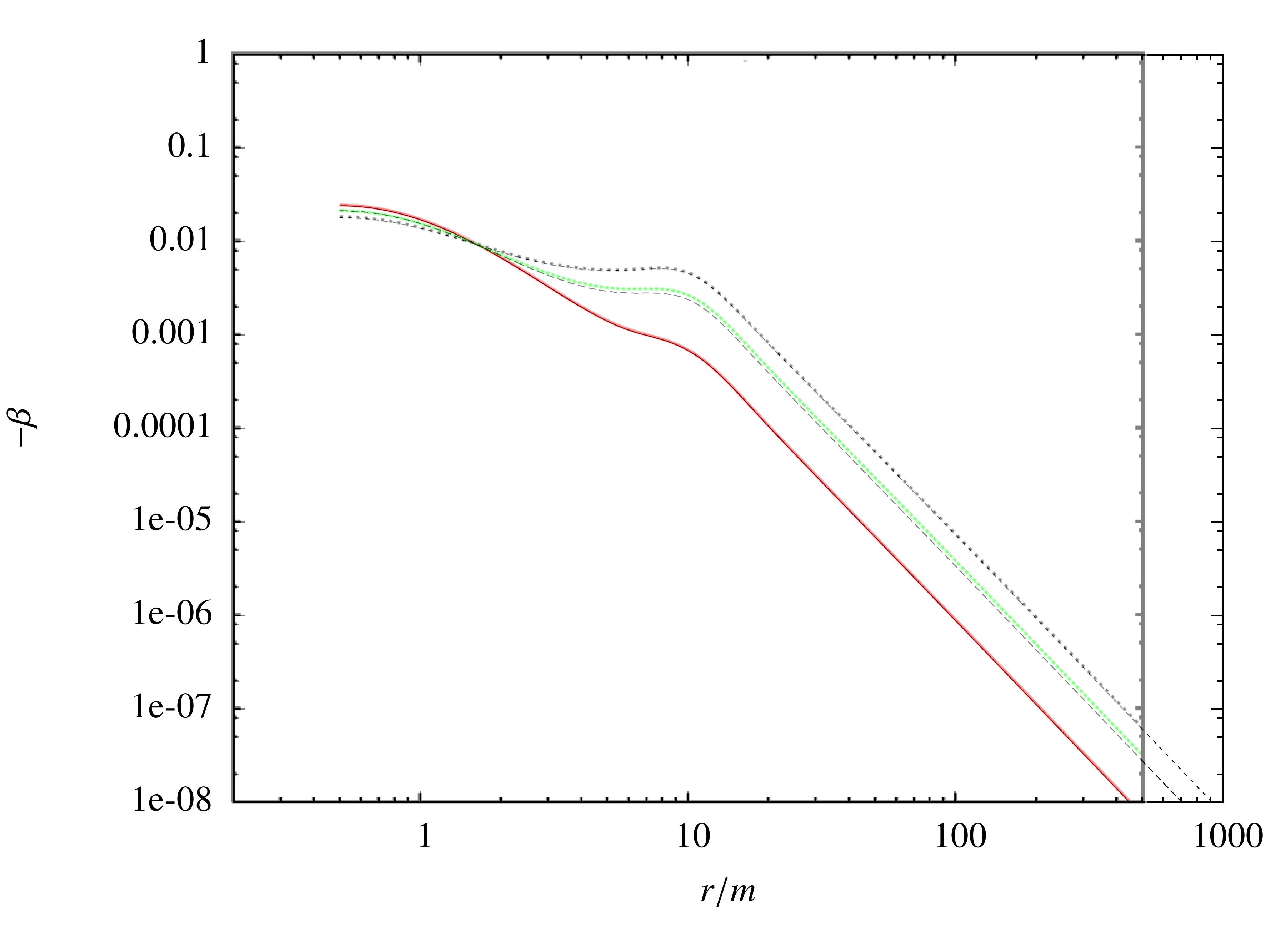}
\caption{\label{fig:por_b}Functions $\phi$ and $-\beta$ in the equatorial
plane for the configuration with the spin parameter of the black hole
$a=0.1$, $r_{1}=5.0$, $r_{2}=20.0$, for the cases $M_{\ast}/m=0.093$,
$\rho_{\mathrm{max}}=0.445\times10^{-4}$ (solid line), $M_{\ast}/m=0.405$,
$\rho_{\mathrm{max}}=1.704\times10^{-4}$ (broken line) and $M_{\ast}/m=0.835$,
$\rho_{\mathrm{max}}=3.083\times10^{-4}$ (dashed-dotted line). Inset
in the two figures (with the abscissa extending up to $r/m=500$)
are the corresponding diagrams taken from \cite{MSH}.}
\end{figure}

We do several calculations using the same parameters and the same
rotation curve\textemdash defined by the constancy of the angular
momentum density $j_{\mathrm{Sh}}=hu_{\varphi}$\textemdash as in
\cite{MSH}. Solutions have been found for parameter sets (spin parameter
$a$, inner $r_{1}$ and outer $r_{2}$ coordinate radii, respectively):
(i) (0.9, 1.5, 20) with $M_{\ast}=0.043,0.4,0.8$ and (ii) (0.1, 5,
20) with $M_{\ast}=0.093,0.405,0.835$. The quantity $M_{\ast}$ defined
as in \cite{MSH} is a kind of a baryonic mass of the torus. Figures
\ref{fig:por_a} and \ref{fig:por_b} portray two families of solutions
of two metric functions $\phi$ and $\beta$ in the plane $\theta=\pi/2$.
The results presented in Figs.\ \ref{fig:por_a} and \ref{fig:por_b}
agree quite well with the results presented in Fig.\ 4 of \cite{MSH}.
There are two exceptions\textemdash the plots of $\phi$ and $-\beta$
($M_{\ast}=0.405$) in Fig.\ \ref{fig:por_b} reveal a difference.

Our solutions presented in Figs.\ \ref{fig:por_a} and \ref{fig:por_b}
were computed on the numerical grids characterized by $\Delta r=r_{\mathrm{s}}/50$,
$f=1.01$, $N_{r}=802$, $N_{\theta}=102$. This yields $r_{\infty}\approx5800r_{\mathrm{s}}$.
The grid nodes were distributed in the angular direction according
to Eq.\ (\ref{eqtheta}).

\section{The coefficient $\kappa$ in the Keplerian rotation law of \cite{MM}}

The rotation law (\ref{momentum}) seems to have two free parameters:
$\kappa$ and $w$. The parameter $\delta$ is dictated by the Newtonian
limit 
\[
\Omega_{0}=\frac{w}{\varpi^{\frac{2}{1-\delta}}},
\]
and for the Keplerian rotation one has to choose $\delta=-1/3$. Then
the formula $\kappa=(1-3\delta)/(1+\delta)$, required by the first
post-Newtonian expansion, yields $\kappa=3$ and 
\begin{equation}
j(\Omega)\equiv\left(-3\,\Omega+w^{-4/3}\Omega^{1/3}\right)^{-1}.\label{momentum1}
\end{equation}

Numerical results, reported below, suggest that in fact the general-relativistic
Keplerian rotation is best described by (\ref{momentum1}). Clearly
the parameter $w$ is not free\textemdash its value constitutes a
part of a solution.

Solutions presented in this and the following sections were computed
on the numerical grids characterized by $\Delta r=r_{\mathrm{s}}/50$,
$f=1.01$, $N_{r}=802$, $N_{\theta}=102$. The grid nodes were distributed
in the angular direction according to Eq.\ (\ref{eqcos}).

\begin{table}[b]
\caption{\label{table1} Rotation around a spinless black hole of the mass
$m=1$. The first 2 columns give the innermost and outermost radii
of tori, the third describes maximal values of the baryonic mass density,
the fourth gives the disk's mass (with $\kappa=3$) and the final
two columns show the minimal and maximal of values of $\kappa$ for
which solutions were found. }

\begin{ruledtabular}
\begin{tabular}{cccccc}
$r_{1}$  & $r_{2}$  & $\rho_{\mathrm{max}}$  & $m_{\mathrm{T}}$  & $\kappa_{\mathrm{min}}$  & $\kappa_{\mathrm{max}}$ \\
\hline 
$5.04$  & $20.1$  & $0.2\times10^{-4}$  & $4.5\times10^{-3}$  & $3.0$  & $4.9$ \\
$5.04$  & $20.1$  & $1.0\times10^{-4}$  & $7.8\times10^{-2}$  & $2.9$  & $4.8$ \\
$5.04$  & $20.1$  & $5.0\times10^{-4}$  & $1.1$  & $1.5$  & $4.0$ \\
$5.04$  & $10.1$  & $0.2\times10^{-3}$  & $7.2\times10^{-3}$  & $3.0$  & $5.2$ \\
$5.04$  & $10.1$  & $1.0\times10^{-3}$  & $1.1\times10^{-1}$  & $2.7$  & $5.0$ \\
$5.04$  & $10.1$  & $4.0\times10^{-3}$  & $1.0$  & $0.8$  & $3.9$ \\
\end{tabular}
\end{ruledtabular}

\end{table}

We performed a large number of numerical calculations for spin-less
central black holes in order to find out whether $\kappa=3$ is the
best choice for the Keplerian rotation. They are shown in Table \ref{table1}.
From these and other calculations, we infer that with the increase
of mass, while keeping the inner boundary close to the innermost stable
circular orbit (ISCO), the interval $(\kappa_{\mathrm{min}},\kappa_{\mathrm{max}})$
for which we were able to find solutions, shifts downwards: lower
and upper bounds for $\kappa$ go down. We can conclude also, that
for the general-relativistic Keplerian rotation with $\delta=-1/3$,
the choice of $\kappa=3$ is the safe choice, that always gives solutions.
Furthermore, calculations with some other pairs of values $\delta,\kappa=(1-3\delta)/(1+\delta)$
always yielded solutions (see next section). That might mean that
$w$ is the only parameter in the rotation law (\ref{momentum1})
that cannot be defined \textit{a priori}.

We investigated also systems with spinning black holes and tori rotating
according to (\ref{momentum1}). We have always found solutions for
corotation ($a>0$ and $\Omega>0$), assuming $\kappa=3$. The situation
is different\textemdash solutions with small masses of the tori do
not necessarily exist for $\kappa=3$\textemdash in the case of counterrotating
systems. In fact, the value of $\kappa$, for which solutions with
light tori exist, can be distinctly different from 3. We have found
a mass gap for counterrotating tori ($\Omega>0$ and $a<0$), with
$\kappa=3$ and $a\in[-0.1,-0.9]$\textemdash tori may exist only
if their mass is larger than a particular mass threshold. The existence
of this mass threshold is interesting from the mathematical point
of view, since it means that the space-time geometry produced by counterrotating
disks with the above specified rotation law does not tend to the Kerr
geometry; solutions seem to disappear when masses of toroids become
too small. On the other hand it appears that this feature is not physical
for the rotation curve (\ref{momentum})\textemdash we have found
that the gap disappears for other values of $\kappa$. Table \ref{table2}
shows such values of $\kappa$ for a selection of spins of central
black holes.

\begin{table}
\caption{\label{table2}The spin of central black holes is given in the first
row. The second row displays values of the parameter $\kappa$ for
which the mass gap disappears. In all examples below the inner and
outer radii are $r_{1}=8$ and $r_{2}=20$, respectively}

\begin{ruledtabular}
\begin{tabular}{cccccc}
$a$  & $-0.9$  & $-0.8$  & $-0.7$  & $-0.6$  & $-0.5$ \\
$\kappa$  & 3.91  & 3.81  & 3.71  & 3.61  & 3.512 \\
$a$  & $-0.4$  & $-0.3$  & $-0.2$  & $-0.1$  & $-0.05$ \\
$\kappa$  & 3.409  & 3.308  & 3.206  & 3.103  & 3.052 \\
\end{tabular}
\end{ruledtabular}

\end{table}

\section{Characteristics of tori rotating around spinless black holes}

We investigated solutions corresponding to the rotation law (\ref{momentum})
for $\delta\in[-0.95,0]$ and $\kappa$ given by the formula $\kappa=(1-3\delta)/(1+\delta)$,
or in its close vicinity. Below we present the following cases: 
\begin{enumerate}
\item [(i)] constant angular momentum density $j$: $\delta=0,\kappa=1$; 
\item [(ii)] the Keplerian rotation $\delta=-1/3,\kappa=3$; 
\item [(iii)] $\delta=-0.8,\kappa=17$; this gives a linear velocity that is slowly
changing across the symmetry plane of the torus; 
\item [(iv)] $\delta=-0.95,\kappa=77$; this gives a linear velocity that varies
very little on the intersection of the symmetry plane with the torus. 
\end{enumerate}
\begin{figure}[t]
\includegraphics[width=1\columnwidth]{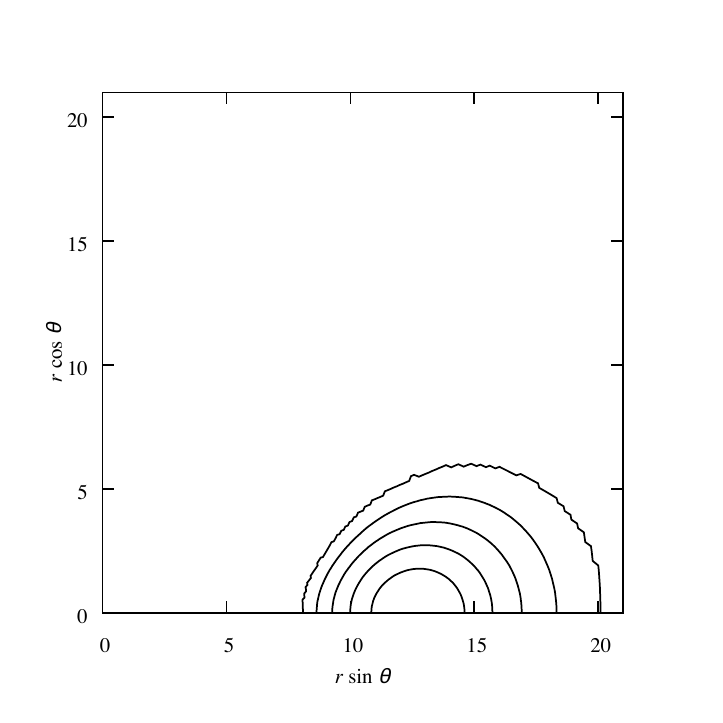} \caption{\label{fig:t01} Isolines of constant enthalpy $h$, for non-spinning
black hole and the rotation law (\ref{momentum}) with $\delta=0$
and $\kappa=1$. The black hole mass $M_{\mathrm{BH}}=1.0567$, $\rho_{\mathrm{max}}=3.45\times10^{-4}$
and $h_{\mathrm{max}}=2.67\times10^{-2}$.}
\end{figure}
\begin{figure}[!bh]
 \includegraphics[width=1\columnwidth]{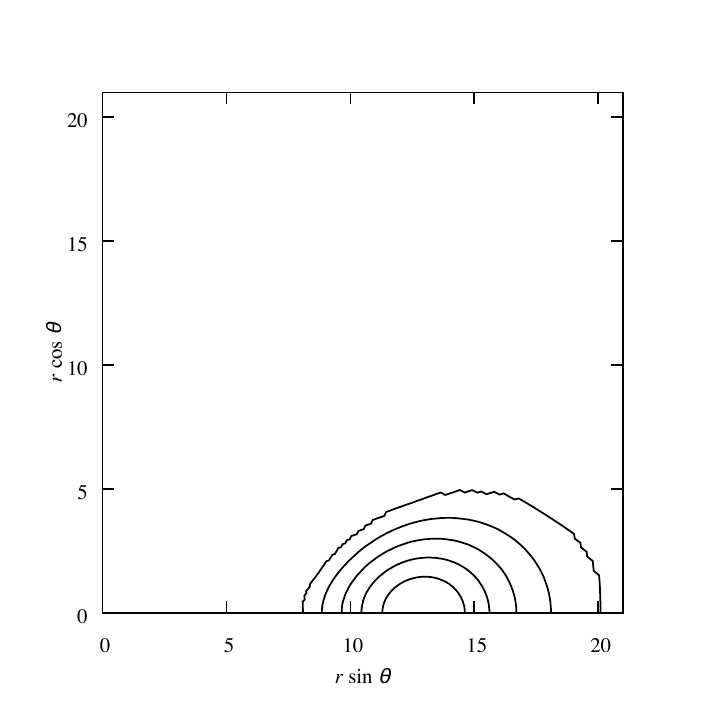} \caption{\label{fig:tK3} Isolines of constant enthalpy $h$, for nonspinning
black hole and the rotation law (\ref{momentum1}). The black hole
mass $M_{\mathrm{BH}}=1.0562$, $\rho_{\mathrm{max}}=4.60\times10^{-4}$
and $h_{\mathrm{max}}=2.31\times10^{-2}$. }
\end{figure}

\begin{figure}[th]
\includegraphics[width=1\columnwidth]{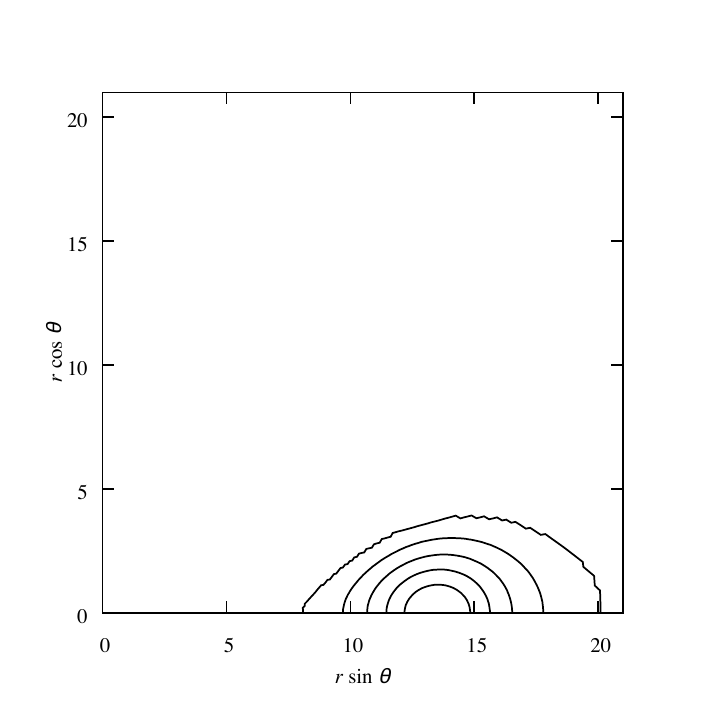} \caption{\label{fig:t-08_17} Isolines of constant enthalpy $h$, for non-spinning
black hole and the rotation law (\ref{momentum}) with $\delta=-0.8$
and $\kappa=17$. The black hole mass $M_{\mathrm{BH}}=1.0542$, $\rho_{\mathrm{max}}=6.97\times10^{-4}$
and $h_{\mathrm{max}}=2.01\times10^{-2}$. }
\end{figure}

\begin{figure}[!b]
\includegraphics[width=1\columnwidth]{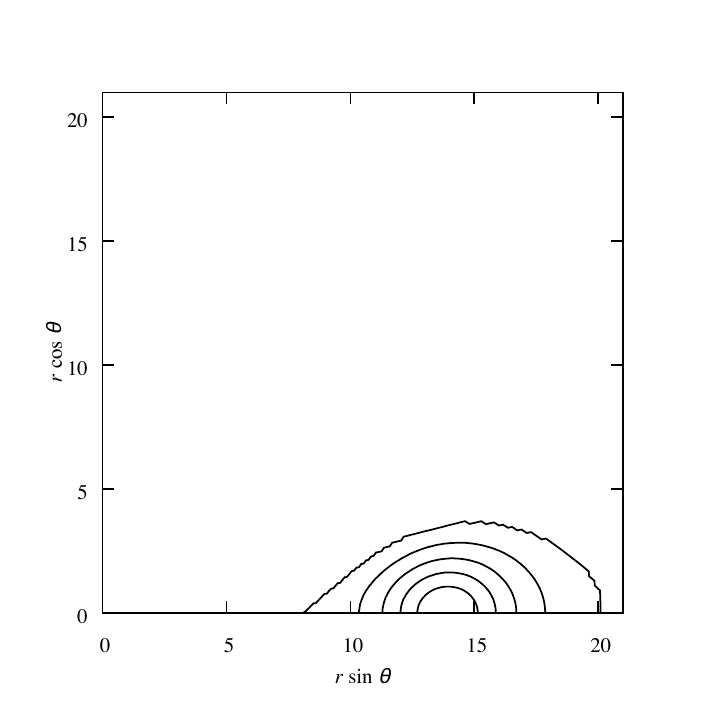} \caption{\label{fig:t-095_77} Isolines of constant enthalpy $h$, for nonspinning
black hole and the rotation law (\ref{momentum}) with $\delta=-0.95$
and $\kappa=77$. The black hole mass $M_{\mathrm{BH}}=1.0531$, $\rho_{\mathrm{max}}=7.93\times10^{-4}$
and $h_{\mathrm{max}}=1.92\times10^{-2}$.}
\end{figure}

The case $\delta=-1$ corresponds to the constant linear velocity,
but it requires a separate numerical implementation of the rotation
law \cite{KMM}.

\begin{figure}[t]
\includegraphics[width=1\columnwidth]{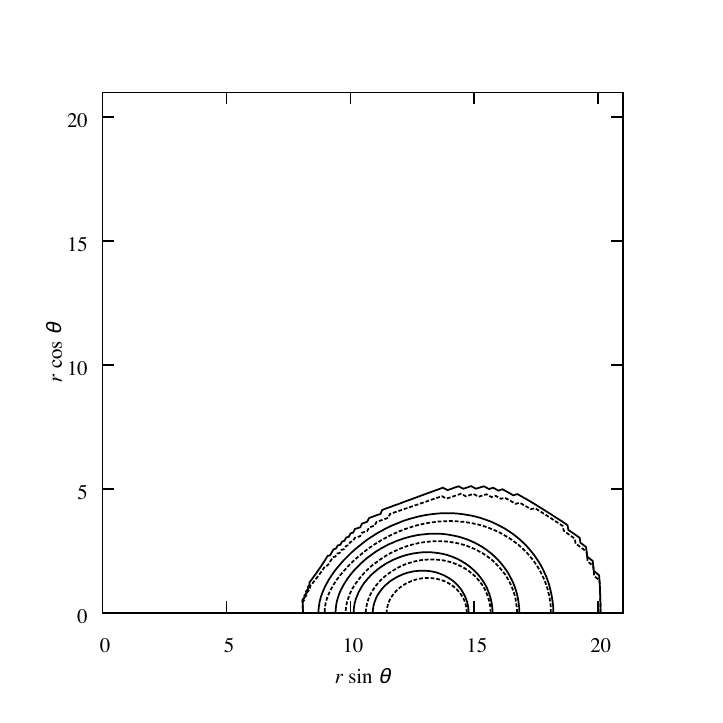} \caption{\label{fig:tK} Isolines of constant enthalpy $h$, for nonspinning
black holes and the rotation law (\ref{momentum1}) with $\delta=-1/3$
for $\kappa=2.5$ (broken lines) and $\kappa=3.5$ (solid lines).
The black hole masses are $M_{\mathrm{BH}}=1.0557$ ($\kappa=2.5$)
and $M_{\mathrm{BH}}=1.0566$ ($\kappa=3.5$). The maximal mass densities
and maximal enthalpy are $\rho_{\mathrm{max}}=4.85\times10^{-4}$
and $h_{\mathrm{max}}=2.24\times10^{-2}$, and $\rho_{\mathrm{max}}=4.37\times10^{-4}$
and $h_{\mathrm{max}}=2.38\times10^{-2}$, for $\kappa=2.5$ and $\kappa=3.5$
respectively.}
\end{figure}

Solutions depicted in Figs.\ \ref{fig:t01}\textendash \ref{fig:tK}
were computed assuming the same asymptotic mass $M_{\mathrm{ADM}}=1.7288$,
spinless black holes and the innermost and outermost coordinate radii
$r_{\mathrm{1}}=8.079$ and $r_{\mathrm{2}}=20.09$, respectively.
The masses of black holes appear to be very similar, which means that
tori masses are essentially the same, $m_{\mathrm{T}}=M_{\mathrm{ADM}}-M_{\mathrm{BH}}\approx0.67$.
The corresponding circumferential radii are similar, $r_{\mathrm{C}}(r_{1})\in(9.64,9.69)$
and $r_{\mathrm{C}}(r_{2})\in(21.99,22.02)$. The 4 inner isolines
in Figs.\ \ref{fig:t01}\textendash \ref{fig:t-095_77} are defined
by the following formula $h_{i}=0.2(4-i)(h_{\mathrm{max}}-1)+1$ for
$i=0,1,2,3$, while the outermost isoline corresponds to $h=1.00000001$.
In Fig.\ \ref{fig:tK} the isolines are defined in the same way,
but $h_{\mathrm{max}}$ is the maximal enthalpy for the configuration
with $\kappa=2.5$.

There is a number of interesting features that depend on the choice
of the rotation curve. 
\begin{enumerate}
\item [(i)] profiles of tori circularize with the increase of $\delta$ and flatten
with the decrease of $\delta$\textemdash inspect Fig.\ \ref{fig:t01}\textendash \ref{fig:t-095_77}.
Tori area also decreases. On the other hand, keeping $\delta=-1/3$
and changing $\kappa$ induces smaller differences (see Fig.\ \ref{fig:tK}). 
\item [(ii)] the maximal mass density $\rho_{\mathrm{max}}$ increases with the
decrease of $\delta$ (see inscriptions to forthcoming figures). The
increase is by a factor of two between the cases of the constant angular
momentum density and the almost linear rotation. The dependence on
the parameter $\kappa$ for fixed $\delta$ seems to be less pronounced
(see Fig.\ \ref{fig:tK}). 
\item [(iii)] the maximal enthalpy $h_{\mathrm{max}}$ decreases with the decrease
of $\delta$ (see captions of the forthcoming figures). 
\end{enumerate}

\section{New rotation law for spinning black holes}

We shall start with deriving the rotation law for a massless disk
of dust encircling a spinning Kerr black hole in the symmetry plane.
As will be shown below, this is a tedious algebraic calculation. To
shorten notation we temporarily use units with $m=1$. First, the
angular velocity $\Omega$ of fluid particles on geodesic circular
orbits in the equatorial plane of the Kerr metric \eqref{metric}
was found, using \eqref{Omega_def}: 
\begin{equation}
\Omega(r)=\frac{8r^{3/2}}{\left((2r+1)^{2}-a^{2}\right)^{3/2}+8ar^{3/2}}.\label{ome}
\end{equation}
The angular momentum \eqref{j_def} reads 
\begin{equation}
j=-\frac{N}{D},\label{jn}
\end{equation}
where 
\begin{eqnarray}
 D & = & 2r^{5}\left(a^{2}-(2r+1)^{2}\right)^{2}\nonumber \\
    & \times &  [a^{7}-3a^{5}(1-2r)^{2}+a^{3}\left(8r\left(r\left(6r^{2}+20r+3\right)-3\right)+3\right)\nonumber \\
    &  - & a(2r+1)^{2}(4(r-2)r+1)^{2}],\nonumber \\
 N& = & \left(ar^{9/2}\left((2r+1)^{2}-a^{2}\right)^{3/2}+8a^{2}r^{6}\right)\nonumber \\
 & \times &  \left(a^{4}+a^{2}(8(r-1)r-2)-16ar\sqrt{r\left((2r+1)^{2}-a^{2}\right)}\right.\nonumber \\
 & + &  \left.(2r+1)^{4}\right)[a^{4}+a^{2}\left(-8r^{2}+4r-2\right)\nonumber \\
 & - & 16ar\sqrt{r\left((2r+1)^{2}-a^{2}\right)}+(2r+1)^{2}(4(r-2)r+1)].\nonumber \\
\label{j_constit}
\end{eqnarray}

Formulas (\ref{ome}) and (\ref{jn}) give the desired relation $j(\Omega)$
in the parametric form. In order to eliminate $r$, notice that (\ref{ome})
can be written as a quadratic equation 
\[
\left(\frac{1}{\Omega}-a\right)^{2/3}=\frac{1}{4r}\left((1+2r)^{2}-a^{2}\right).
\]
Denoting 
\begin{equation}
\xi=\left(\Omega^{-1}-a\right)^{1/3},\label{xi}
\end{equation}
solving it for $r$ and substituting the result into the formula for
$j(r)$ would give, in principle, the desired result. Unfortunately,
it takes form of nested radicals composed of a large number ($\sim10^{3}$)
of terms. This is a difficult problem for nonrational values of $a$
\cite{NestedRoot}; computer algebraic systems do not help in simplifying
such formulae. On the other hand, we observe that the formula greatly
simplifies for rational $a$, e.g., $a=-2/3$. The presence of multiple
repetitions of roots of the numerator and the denominator strongly
suggests the existence of a much simpler expression. A simplification
of the formula for $j(\Omega)$ proceeds as follows. We define 
\begin{equation}
X=\sqrt{a^{2}-2\xi^{2}+\xi^{4}}
\end{equation}
and 
\begin{equation}
Z=\sqrt{1-a^{2}+2X(1+X-\xi^{2})}.
\end{equation}
One can write the formula (\ref{jn}) in the form 
\begin{equation}
j(\xi)=-\frac{N_{1}N_{2}N_{3}}{D_{1}D_{2}D_{3}D_{4}},\label{j11}
\end{equation}
where 
\begin{eqnarray}
N_{1} & = & a+\xi^{3},\nonumber \\
N_{2} & = & \xi\left(\xi^{2}-3\right)Z^{2}+2a\left(1+X-\xi^{2}\right)Z,\nonumber \\
N_{3} & = & \left(\xi^{4}+a^{2}\right)Z^{2}+2a\xi\left(1+X-\xi^{2}\right)Z,\label{N2}
\end{eqnarray}
\begin{eqnarray}
D_{1} & = & \xi^{3}, \ \ \ D_{2} = \xi^{2}\left(\xi^{2}-3\right)^{2}-4a^{2},\nonumber \ \ \ D_{3} = \xi^{2}-1-X,\nonumber \\
D_{4} & = & 1+3X-\left(9+8X\right)\xi^{2}+4\left(3+X\right)\xi^{4}-4\xi^{6}\nonumber \\
 & + & a^{2}(3+X-3\xi^{2}).
\end{eqnarray}
One can easily check that $D_{3}D_{4}=-Z^{4}$, so that we get 
\begin{equation}
j(\xi)=\frac{\left(a+\xi^{3}\right)N_{2}N_{3}}{D_{2}\xi^{3}Z^{4}}.\label{j2}
\end{equation}
Now we rewrite Eq.\ (\ref{N2}) as follows: 
\[
2a\left(1+X-\xi^{2}\right)Z=N_{2}-\xi\left(\xi^{2}-3\right)Z^{2};
\]
by squaring the above equation we get rid of the square root in $Z$.
We solve the squared equation with respect to $N_{2}$. The physical
root reads 
\begin{equation}
N_{2}=\left[\xi\left(\xi^{2}-3\right)-2a\right]Z^{2}.\label{N2a}
\end{equation}
We proceed similarly with the factor $N_{3}$, obtaining the physical
root: 
\begin{equation}
N_{3}=\left(\xi^{4}-2a\xi+a^{2}\right)Z^{2}.\label{N3}
\end{equation}
On the other hand 
\begin{equation}
D_{2}=\left[\xi\left(\xi^{2}-3\right)-2a\right]\left[\xi\left(\xi^{2}-3\right)+2a\right].\label{D2}
\end{equation}
After inserting (\ref{N2a}), (\ref{N3}) and (\ref{D2}) into (\ref{j2})
one ends up with 
\begin{equation}
j\left(\xi\right)=\frac{\left(a+\xi^{3}\right)\left(\xi^{4}-2a\xi+a^{2}\right)}{\xi^{3}\left[\xi(\xi^{2}-3)+2a\right]}.\label{j(xi)}
\end{equation}
Finally, using $\left(\ref{xi}\right)$ one obtains the following
formula: 
\begin{equation}
j\left(\Omega\right)=-\frac{1}{2}\frac{d}{d\Omega}\ln\left\{ 1-\left[a^{2}\Omega^{2}+3\Omega^{\frac{2}{3}}\left(1-a\Omega\right)^{\frac{4}{3}}\right]\right\} ,\label{j(O)}
\end{equation}
which can be easily generalized to the case with $m=w^{2}\neq1$,
by the substitution $\Omega\rightarrow w^{2}\Omega$, $a\rightarrow a/w^{2}$,
$j\rightarrow j/w^{2}$. We get 
\begin{equation}
j\left(\Omega\right)=-\frac{1}{2}\frac{d}{d\Omega}\ln\left\{ 1-\left[a^{2}\Omega^{2}+3w^{\frac{4}{3}}\Omega^{\frac{2}{3}}\left(1-a\Omega\right)^{\frac{4}{3}}\right]\right\} .\label{j(O)-mass}
\end{equation}
We shall write this expression as 
\begin{equation}
j\left(\Omega\right)=-\frac{3}{2\kappa}\frac{d}{d\Omega}\ln\left\{ 1-\frac{\kappa}{3}\left[a^{2}\Omega^{2}+3w^{\frac{4}{3}}\Omega^{\frac{2}{3}}\left(1-a\Omega\right)^{\frac{4}{3}}\right]\right\} .\label{j(O)-1}
\end{equation}
It is easy to see that for $a=0$ and $\kappa=3$ we recover the spinless
rotation curve (\ref{momentum1}).

\noindent The present numerical code was also adapted to the new rotation
curve. Section IV.A.5 reports results on its numerical convergence
and the second example in Sec. VIII also uses the new law (\ref{j(O)-mass}).

\textit{We conjecture that the rotation law (\ref{j(O)-1}) holds
not only for massless disks of dust, but also for massive toroids,
and with the parameter $\kappa=3$.} This hypothesis is supported
by partial numerical results.

\section{Testing the Poincar\'{e}-Wavre property in general-relativistic
Keplerian toroids}

The general relativistic and Newtonian rotations of cylindrically
symmetric tori circulating around a symmetry axis differ fundamentally
in one important aspect. In the Newtonian case the angular velocity
$\Omega$ is a function of the geometric distance from the rotation
axis only. This is a classical result of Poincar\'{e}-Wavre.
Rephrasing the same in the terminology of \cite{Kozlowski}, the von
Zeipel cylinder of constant $\Omega$ lies at a constant geometric
distance from the symmetry axis. In contrast to that, in general relativity
the angular velocity of a particle of fluid within a torus becomes
a function both of its circumferential distance from the rotation
axis and the distance from the plane of the symmetry. Different points
lying on the von Zeipel cylinder are not equidistant from the symmetry
axis. This might manifest observationally through widening of spectral
lines of radiation emitted by sources lying on the von Zeipel cylinder.
The same would be true about characteristic radiation lines of sources
located on an circumferentially-equidistant cylinder, as hinted in
the 1PN analysis of spinless black holes \cite{JMMP,MM,MMP2015,KMMPX}.

We shall discuss two solutions. One of them is a system obeying the
rotation law (\ref{momentum1}) with a spin-less black hole. In the
second solution the toroid rotates according to (\ref{j(O)-mass})
(with $a=0.9$), while the black hole has a large spin parameter $a=0.9$.
We show the dependence of the angular velocity along a cylinder having
a fixed value of the circumferential radius $r_{\mathrm{C}}$.

\begin{figure}[t]
\includegraphics[width=1\columnwidth]{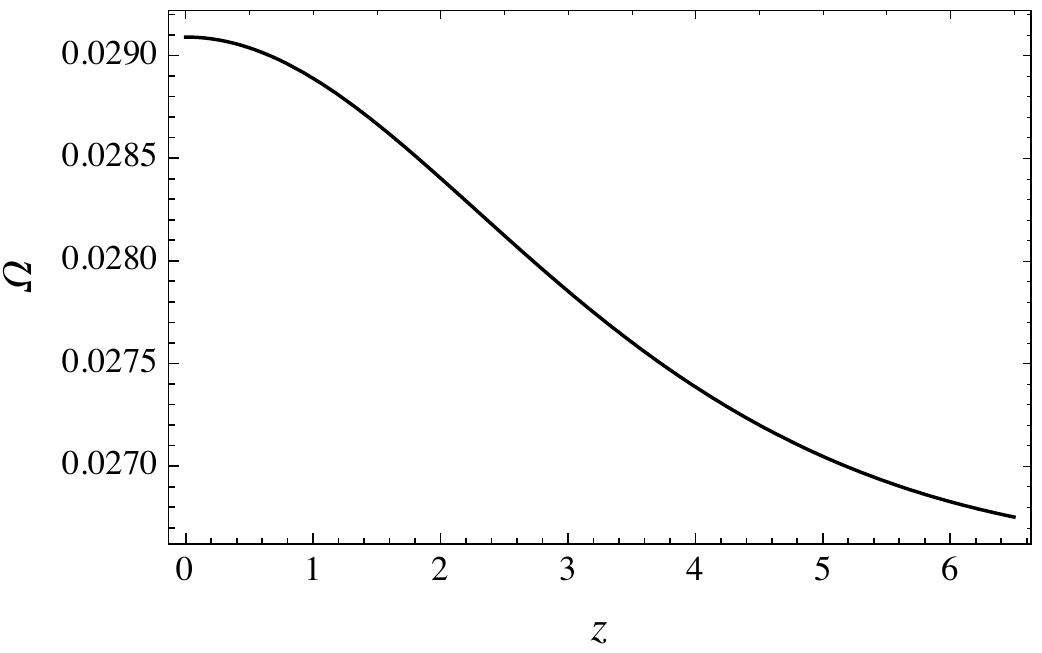} \caption{\label{fig:omega_z} Nonspinning black hole and the rotation law (\ref{momentum1}).
The innermost and outermost coordinate radii are $r_{\mathrm{1}}=5.037$
and $r_{\mathrm{2}}=20.09$, respectively. The coordinate $z=r\cos\theta$.
The maximal mass density $\rho_{\mathrm{max}}=7\times10^{-4}$.}
\end{figure}

Figure \ref{fig:omega_z} corresponds to the non-spinning black hole
characterized by $M_{\mathrm{BH}}=1.29$. The asymptotic mass of the
system reads $M_{\mathrm{ADM}}=3.74$. The circumferential radii of
the innermost and outermost points on the torus are $r_{\mathrm{C}}(r_{\mathrm{1}})=8.01$
and $r_{\mathrm{C}}(r_{\mathrm{2}})=24.36$. The largest relative
change of the angular velocity along a cylinder having the circumferential
radius $r_{\mathrm{C}}=14.29$ exceeds 9\%.

\begin{figure}[t]
\includegraphics[width=1\columnwidth]{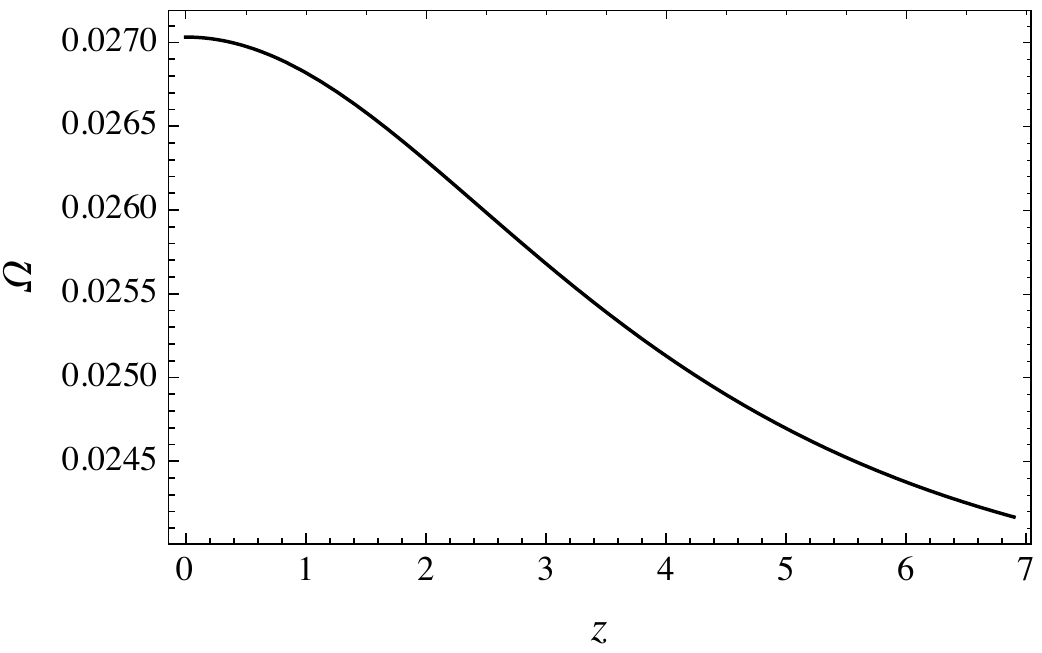} \caption{\label{fig:omega_z2} Spinning black hole and the rotation law (\ref{j(O)-1}).
The innermost and outermost coordinate radii are $r_{\mathrm{1}}=5.027$
and $r_{\mathrm{2}}=20.08$, respectively. The coordinate $z=r\cos\theta$.
The maximal mass density $\rho_{\mathrm{max}}=8\times10^{-4}$.}
\end{figure}

Somewhat stronger effects are shown for spinning black holes. The
corresponding example is shown in Fig.\ \ref{fig:omega_z2}. The
model is characterized by the black hole mass $M_{\mathrm{BH}}=1.033$
and its spin parameter $a=0.9$. The asymptotic mass of the system
reads $M_{\mathrm{ADM}}=4.074$. The circumferential radii of the
innermost and outermost points on the torus are $r_{\mathrm{C}}(r_{1})=8.05$
and $r_{\mathrm{C}}(r_{2})=24.79$. The largest relative change of
the angular velocity along a cylinder having the circumferential radius
$r_{\mathrm{C}}=15.09$ exceeds 11\% (see Fig.\ \ref{fig:omega_z2}).

In both cases the dependence on height is quite robust, particularly
for the spinning black hole. The angular velocity drops down from
the maximal value on the equator to the minimum value on the edge
of the disk. Admittedly these examples have unrealistically large
masses of tori.

\section{Summary}

We report in this paper results of a successful numerical implementation
of a recent rotation law \cite{MM} in the full Einstein theory. This
is done within the puncture framework as implemented by Shibata \cite{MSH},
but with several modifications.

Our numerical codes are now capable to describe a class of self-gravitating
tori in the regime of strong gravity for those general-relativistic
rotation curves that in the Newtonian limit yield angular velocities
$\Omega\propto1/\varpi^{\lambda}$, with $0\le\lambda<1$. This class
includes stationary disks in tight accretion systems with central
(spinless or spinning) black holes, and it contains polytropic tori
with the general relativistic Keplerian rotation, almost constant
linear velocity and the constant angular momentum density.

Our numerical codes successfully passed several numerical tests. We
recovered many of solutions found in \cite{MSH}. Our results agree
well in the post-Newtonian regime with the earlier post-Newtonian
analysis (see \cite{KMMPX,KMM,JMMP}), but the related material will
be published elsewhere.

Section VI shows profiles of several tori. It is interesting that
they differ morphologically\textemdash they flatten when the rotation
curve approaches the constant linear velocity, and become ``fatter''
for Keplerian rotation and in the case of constant angular momentum.

We have found in Sec.\ VII a new general-relativistic Keplerian rotation
law, that generalizes the former one \cite{MM}; it is expected to
be more effective in the case of spinning central black holes. The
present numerical code successfully operates\textemdash after adaptation\textemdash with
the new rotation curve (see some results in Secs.\ IV.A.5 and VIII),
but the investigation is still under way.

It is known that highly relativistic tori in almost Keplerian rotation
can be created in the merger of compact binaries consisting of pairs
of black holes or neutron stars \cite{NSkeplerian,KR,ECGKK,SFHKKST,Pan_Ton_Rez,Lovelace},
associated with the emission of gravitational waves \cite{GW17082017}.
Their formation seems to be important for understanding features of
the electromagnetic radiation associated with these mergers. Our codes
allow a quick manufacturing of such polytropic tori\textemdash this
is a question of tens of minutes or at most hours on a standard desktop
computer. They could be used as initial data for dynamic systems supplied
with all required physics.

Rotating tori can exist in some active galactic nuclei. They might
reveal some general-relativistic effects due to the phenomena discussed
here. We have shown that the general-relativistic angular velocity
is height-dependent (\ref{fig:omega_z}); that might cause widening
of spectral lines of radiation emitted by the disk (via Doppler effect)
and obscure the interpretation of the rotation, especially for tight
AGN systems.

Rotation curves (\ref{momentum}) might be used for the description
of rotating stars, after appropriate modification in the vicinity
of the symmetry axis.
\begin{acknowledgments}
This research was carried out with the supercomputer ``Deszno''
purchased thanks to the financial support of the European Regional
Development Fund in the framework of the Polish Innovation Economy
Operational Program (Contract no.\ POIG.\ 02.01.00-12-023/08). P.\ M.
acknowledges the financial support of the Narodowe Centrum Nauki Grant
No. DEC-2012/06/A/ST2/00397. M.\ P. acknowledges partial support
from the Grant No, K/DSC/004356.
\end{acknowledgments}

\end{document}